\documentclass{JFM-FLM_Au}

\usepackage{amsmath,amssymb}
\usepackage{graphicx}
\usepackage{booktabs,array}
\usepackage{float}
\usepackage{placeins}
\usepackage{xcolor}
\usepackage{url}
\usepackage{hyperref}
\usepackage{etoolbox}
\usepackage{CJKutf8}
\newcommand{\chinesename}[1]{\begin{CJK*}{UTF8}{gbsn}\fontseries{bx}\selectfont#1\end{CJK*}}

\makeatletter
\def\oddabsfooterflag{\hbox to \textwidth{\hfill{\cppagefont 0 X0-\thepage}}}
\def\evenabsfooterflag{\hbox to \textwidth{{\cppagefont 0 X0-\thepage}\hfill}}
\def\pagelimitfooter{\hbox to \textwidth{{\cppagefont 0 X0-\thepage}\hfill}}
\let\manuscript@titlepage\ps@titlepage
\def\ps@titlepage{\manuscript@titlepage\let\@oddhead\@empty\let\@evenhead\@empty}
\patchcmd{\@maketitle}{\vspace*{10\p@}\addvspace{4.6pc}}{\vspace*{10\p@}}{}{\PackageError{manuscript}{Title spacing patch failed}{Check the authoring class.}}
\makeatother

\newcommand{\ftlesoftware}{\texttt{SCP-FTLE}}
\newcommand{\fslesoftware}{\texttt{SCP-FSLE}}
\newcommand{\software}{\textsc{SCP-FTLE} and \textsc{SCP-FSLE}}

\lefttitle{Wei}
\righttitle{High-performance FTLE and FSLE computations for CPUs and GPUs}

\title{\software: High-performance finite-time and finite-size Lyapunov-exponent computations for CPUs and GPUs}

\author{Huanxia Wei (\chinesename{韦欢夏})\aff{1}, Mengqi Zhang (\chinesename{张蒙齐})\aff{2}}

\affiliation{\aff{1}Department of Mechanical and Aerospace Engineering, The University of Manchester, Oxford Road, Manchester M13 9PL, UK\\
\aff{2}Department of Mathematics, City University of Hong Kong, 83 Tat Chee Avenue, Kowloon Tong, Hong Kong}
\corresau{Huanxia Wei, \email{huanxia.wei@manchester.ac.uk}}

\begin{document}

\maketitle

\begin{abstract}
Finite-time and finite-size Lyapunov exponents (FTLE and FSLE) characterise material stretching and transport barriers over prescribed time intervals and separation scales, but their evaluation from three-dimensional velocity sequences requires large computational costs. We introduce \ftlesoftware{} and \fslesoftware{}, two modules of \textsc{StreamCenter+} for CPU- and GPU-accelerated Lagrangian analysis. The modules support structured and native unstructured meshes, selectable integration and interpolation schemes, and graphical and command-line workflows. Rolling snapshot reuse, memory-bounded FTLE batching, device-resident trajectory state and active-set FSLE advancement reduce data movement and redundant computation, achieving the state-of-the-art (SOTA) computational performance among peer packages. For a billion-seed benchmark, strong-scaling efficiencies on 32 GPUs reach 86.3\% for FTLE and 95.2\% for FSLE. 
FTLE is verified with published steady and unsteady Arnold--Beltrami--Childress (ABC) flows and stationary concentrated vortex model (SCVM), while FSLE is verified in linear stretching and rotational shear flows. 
As examples, applications reveal the spanwise reorganisation of stretching layers in a shear-thinning lid-driven cavity, and resolve three-dimensional stretching structures associated with contrasting tracer dispersion around an urban building canopy at NUS campus. These results demonstrate a scalable workflow in large-scale Lagrangian diagnostics and analysis of flow data.
\end{abstract}

\section{Introduction}
\label{sec:introduction}

Finite-time Lyapunov exponents (FTLE) and finite-size Lyapunov exponents (FSLE) are Lagrangian diagnostics for quantifying particle separation and analysing material transport and mixing in unsteady flows \citep{Shadden2005,Aurell1997}. They characterise particle separation over prescribed time intervals and separation scales, respectively. Within the broader framework of Lagrangian coherent structures (LCS), such diagnostics help identify material structures associated with strong attraction, repulsion and persistent fluid coherence \citep{Haller2015}. Applications of FTLE, FSLE and the LCS framework span geophysical transport and oceanic mixing \citep{Samelson2013,dOvidio2004}, transport in space plasmas \citep{Rempel2023}, and material organisation in canonical wall-bounded flows \citep{Bettencourt2013,Rist2023}.

The FTLE measures the maximum average exponential separation rate of initially neighbouring trajectories over a fixed integration interval. Forward- and backward-time FTLE ridges are commonly used as indicators of repelling and attracting material structures, respectively \citep{Haller2001,FarazmandHaller2013}. The FSLE, introduced to characterise predictability beyond infinitesimal perturbations \citep{Aurell1997}, instead uses the time required for trajectories to separate between two prescribed distances \citep{Artale1997}. Whereas FTLE compares deformation over a common duration, FSLE compares the rates at which trajectories reach a common separation scale. 

Several existing computational packages have made these methods accessible to fluid-mechanics research, but in a 2-D form. \citet{Onu2015} developed \texttt{LCS Tool}, a \texttt{MATLAB} platform for extracting hyperbolic and elliptic coherent structures from 2-D velocity fields. The \texttt{LCS MATLAB Kit} also provides FTLE calculation and includes an extension to inertial particles \citep{DabiriLCSKit}. \citet{Conti2012} accelerated 2-D FTLE calculations on GPUs and accelerated processing units using OpenCL, demonstrating the benefits of parallel interpolation and trajectory integration. \citet{Sagrista2020} introduced an 2-D interactive framework for exploring FTLE fields and their dependence on integration parameters. 

Other developments address 3-D computation and integration with simulation data. \texttt{TBarrier} brings together a broader collection of transport-barrier methods in Python notebooks \citep{TBarrier2023}, and \texttt{NumbaCS} uses just-in-time compilation to accelerate trajectory integration and coherent-structure analysis \citep{JarvisRoss2025}. Both packages currently use CPU execution without native GPU or multi-node support. \citet{Ameli2014} developed \texttt{FlowTK}, combining \texttt{VTK}-based processing with CUDA kernels for single-GPU execution and integration into \texttt{ParaView}. \citet{Dauch2018} developed GPU-based FTLE evaluation directly from precomputed 3-D smoothed-particle-hydrodynamics (SPH) particle trajectories, thereby avoiding additional trajectory integration during post-processing. However, this strategy cannot be used on velocity fields from general CFD studies. \texttt{UVaFTLE} supports parallel 2-D and 3-D FTLE evaluation on shared-memory CPUs and multiple GPUs within a single node \citep{UVaFTLE2023}. Its inputs include mesh coordinates, connectivity and a precomputed flow map, so trajectory integration must be supplied separately, which is a heavy computational process as the bottleneck of the total workflow. They developed a compatible independent package, \texttt{UVaFlow} \citep{UVaFlowRepository}, for flowmap computation, but in a CPU form. \texttt{FlowVC} provides particle tracking and FTLE evaluation on Cartesian and simplicial meshes through a dedicated binary data format, based on serial-CPU execution \citep{FlowVC2010}. \citet{Lipinski2012} proposed ridge tracking to concentrate FTLE evaluations near coherent-structure surfaces to reduce the approximate computational scaling from $O(\Delta x^{-3})$ to $O(\Delta x^{-2})$ by avoiding evaluations away from the ridges. This is particularly useful for 3D surface extraction, although complete volumetric fields remain necessary when the performing quantitative study of tensile and aggregation strength, rather than focusing on a few representative surfaces. 
These limitations motivate an integrated platform for full 3-D FTLE evaluation with efficient CPU/GPU execution and multi-node scalability.

The situation of FSLE computing packages is similar to FTLE, and even more scarce. The \texttt{AVISO/lagrangian} package computes forward- and backward-time FSLE and FTLE maps from time-dependent horizontal velocity fields in NetCDF format \citep{AVISOLagrangian}. The \texttt{SPASSO} workflow incorporates FSLE through the \texttt{LAMTA} package to support adaptive sampling during oceanographic cruises \citep{SPASSO2019}. These implementations connect Lagrangian diagnostics to observed surface circulation, but their horizontal mapping workflows do not directly provide volumetric analysis of three-component CFD velocity fields. \texttt{VisIt}'s LCS operator also supports FTLE and FSLE calculations within a general-purpose visualisation environment \citep{VisItLCSDocs}. At larger computational scales, \texttt{Aquila-LCS} provides CPU- and GPU-based 3-D FTLE and FSLE calculations and reports particle ensembles reaching billion-level \citep{Lagares2024}. They use first-order Euler advection with trilinear spatial and linear temporal interpolation \citep{AquilaRepository}, and provide no selectable higher-order integration or interpolation schemes, limiting systematic assessment of the accuracy--cost trade-off across numerical orders.

Despite these advances, the simultaneous requirements of numerical flexibility, computational efficiency and compatibility with modern CFD remain difficult to satisfy. Packages developed for planar flows cannot capture fully 3-D separation, while implementations tied to a particular data representation require additional conversion before they can process other simulation outputs. Separating trajectory integration from diagnostic evaluation also complicates consistent control of interpolation, time stepping and boundary treatment. For long 3-D velocity sequences, these limitations are compounded by the cost of repeated data loading, the memory required for particle ensembles and intermediate flow maps, and consideration of MPI structures. An integrated platform must therefore address numerical approaches, data access, parallel execution and optimization together.

\begin{figure}[!t]
  \centering
  \includegraphics[width=\textwidth]{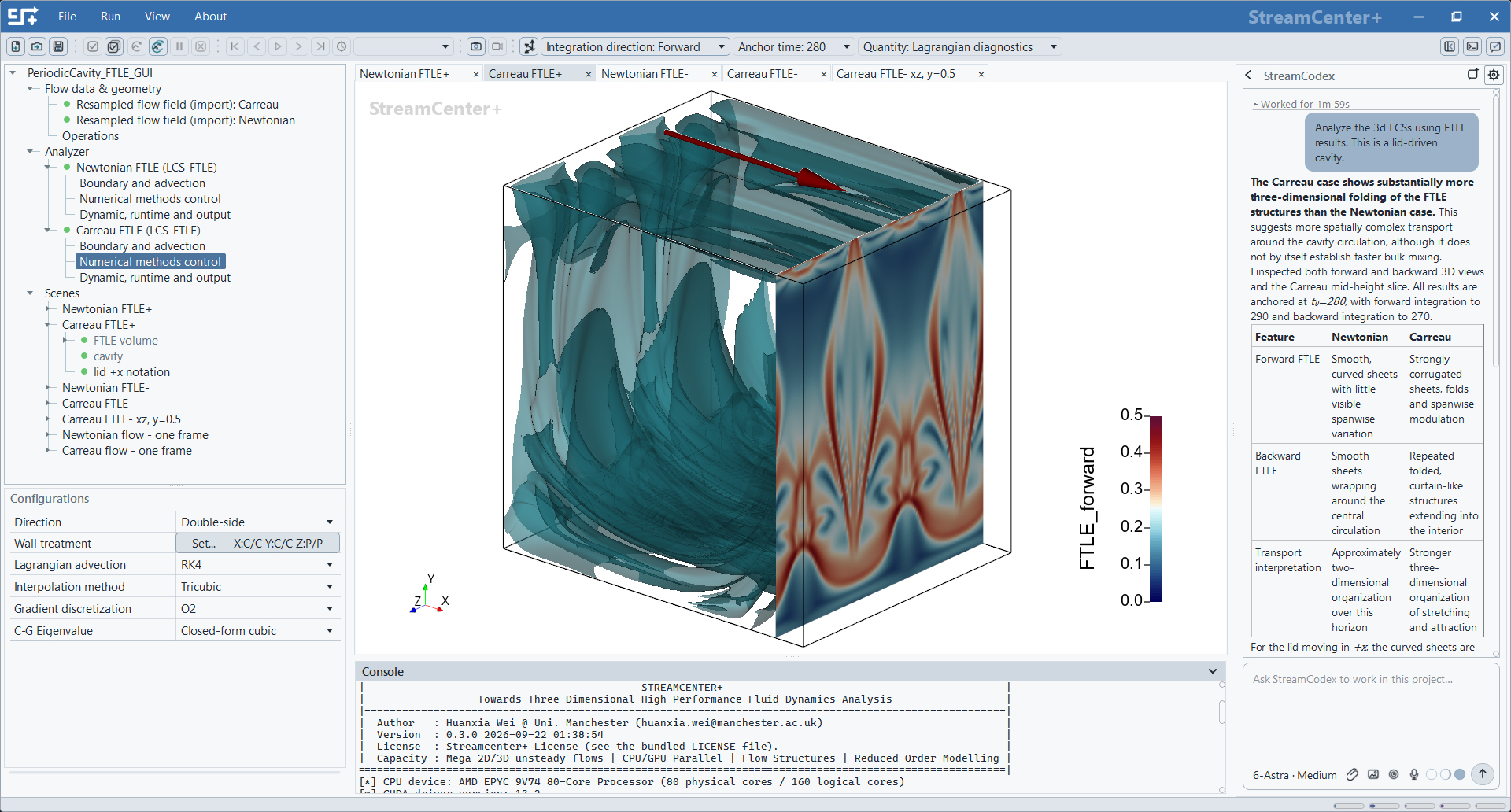}
  \caption{\textsc{StreamCenter+} GUI platform. The presented case is further detailed in \S\ref{sec:application-cavity}.}
  \label{fig:streamcenter-gui}
\end{figure}

This paper presents \ftlesoftware{} and \fslesoftware{}, two dedicated computational modules within our \textsc{StreamCenter+} (SCP) platform (Figure~\ref{fig:streamcenter-gui}) which integrates high-performance flow diagnostics and reduced-order modelling. We integrate multiple interpolation, trajectory-integration and boundary-treatment methods, allowing numerical method choices to be adapted to the resolution and physical characteristics of the input flow. Structured and unstructured meshes are supported, with multicore CPU execution and GPU acceleration for multiple nodes. Readers for common CFD formats and VTK result output connect the calculations to existing simulation and visualisation workflows. A desktop interface supports interactive configuration, while command-line execution enables batch processing on high-performance computing systems. Reuse of velocity snapshots and trajectory state reduces repeated work in FTLE analysis, and event-driven processing avoids unnecessary continuation of completed FSLE trajectories. Memory-bounded data processing supports fully 3-D analysis of velocity datasets containing $>O(10^{10})$ space--time samples. The software provides an extensible computational platform for large-scale FTLE and FSLE analysis.

The remainder of this paper is structured as follows: \S\ref{sec:diagnostics} defines the FTLE and FSLE diagnostics. \S\ref{sec:implementation} describes the numerical methods, execution strategy and computational performance. \S\ref{sec:validation} presents numerical verification, and \S\ref{sec:applications} illustrates flow applications. \S\ref{sec:conclusions} summarises the conclusions.

\section{Lagrangian diagnostics}
\label{sec:diagnostics}

Let $\boldsymbol{u}(\boldsymbol{x},t)$ be a sufficiently regular velocity field and let a trajectory satisfy
\begin{equation}
  \dot{\boldsymbol{x}}=\boldsymbol{u}(\boldsymbol{x},t),
  \qquad \boldsymbol{x}(t_0)=\boldsymbol{x}_0.
  \label{eq:trajectory}
\end{equation}
The flow map $\boldsymbol{\Phi}_{t_0}^{t_0+T}$ sends the initial position $\boldsymbol{x}_0$ to its position after the signed interval $T$. Forward integration uses $T>0$ and backward integration uses $T<0$.

For FTLE, the deformation gradient and right Cauchy--Green tensor are
\begin{equation}
  \boldsymbol{F}_{t_0}^{t_0+T}=\nabla_{\!\boldsymbol{x}_0}
  \boldsymbol{\Phi}_{t_0}^{t_0+T},
  \qquad
  \boldsymbol{C}_{t_0}^{t_0+T}=\boldsymbol{F}^{\mathsf T}\boldsymbol{F}.
  \label{eq:cg}
\end{equation}
With $\lambda_{\max}$ denoting the largest eigenvalue of $\boldsymbol{C}$, the FTLE is
\begin{equation}
  \sigma_{t_0}^{T}(\boldsymbol{x}_0)
  =\frac{1}{|T|}\log\sqrt{\lambda_{\max}
  \left(\boldsymbol{C}_{t_0}^{t_0+T}(\boldsymbol{x}_0)\right)}.
  \label{eq:ftle}
\end{equation}
The solver evaluates the flow map on a user-defined Cartesian seed lattice. Centred finite differences approximate $\boldsymbol{F}$ in the interior, with periodic differencing for periodic domains. The largest eigenvalue of the symmetric $3\times3$ tensor is obtained from the closed-form solution of its characteristic cubic. 
Large forward-time FTLE values indicate strong finite-time separation, while large backward-time values indicate strong convergence in forward physical time. Scalar ridges are therefore candidates for repelling and attracting transport barriers.

For FSLE, let $\delta_0>0$ be an initial separation and $\delta_f>\delta_0$ a target separation, and consider a finite set $\mathcal{N}(\boldsymbol{x}_0)$ of neighbouring seeds placed at distance $\delta_0$ from the central seed. The first separation time is
\begin{equation}
  \tau(\boldsymbol{x}_0)=\inf\left\{|t-t_0|:
  \max_{\boldsymbol{q}\in\mathcal{N}(\boldsymbol{x}_0)}
  \left\|\boldsymbol{\Phi}_{t_0}^{t}(\boldsymbol{q})-
  \boldsymbol{\Phi}_{t_0}^{t}(\boldsymbol{x}_0)\right\|
  \geq \delta_f\right\}.
  \label{eq:tau}
\end{equation}
The FSLE reported at the central seed is
\begin{equation}
  \lambda_{\mathrm{FS}}(\boldsymbol{x}_0)
  =\frac{\log(\delta_f/\delta_0)}{\tau(\boldsymbol{x}_0)}.
  \label{eq:fsle}
\end{equation}
The exponent uses the positive elapsed time, while a separate signed event time preserves whether the branch was integrated forward or backward. 

Neighbour directions could be restricted to the positive and negative coordinate axes or expanded to all normalised sign directions. The axial stencil contains six neighbours in three dimensions and four in two dimensions, while the full stencil contains 26 and eight, respectively. The earliest valid crossing over the stencil determines $\tau$. For a periodic domain, distances use the minimum-image convention, which avoids a false large separation when a pair straddles a periodic face. When a threshold lies between two accepted trajectory states, the crossing time is linearly interpolated from the two separation distances.

\section{Numerical implementation and performance}
\label{sec:implementation}

\subsection{Software capabilities and numerical workflow}
\label{sec:software}

Table~\ref{tab:software} first summarises the capabilities of \ftlesoftware{} and \fslesoftware{}. Time-resolved flow data are read, a diagnostic lattice is constructed over the selected physical region, trajectories are advanced, results are assembled, and the resulting scalar and status fields are written. Regular-grid variants sample a Cartesian velocity representation, whereas native unstructured-mesh variants interpolate directly within the cells of the supplied flow mesh. The diagnostic lattice can differ from the velocity mesh, allowing spatial convergence to be studied independently of the original flow samples. 
Input format, geometry and time ordering are automatically checked before advection.

\begin{table}
  \begin{center}
    \begin{tabular}{p{0.23\textwidth}p{0.67\textwidth}}
      \textbf{Characteristic} & \textbf{Supported options} \\[3pt]
      Diagnostics & Static and dynamic FTLE, first-separation-time FSLE \\
      Dimensionality & 2-D and 3-D velocity sequences \\
      Directions & Forward, backward, or both time directions \\
      Integration & Euler, RK2, RK4, RK6(7) \\
      Spatial sampling & Structured grids: linear, trilinear, cubic Hermite, tricubic and WENO5 interpolation. Native unstructured meshes: cell-based vtk interpolation. \\
      Boundaries & Constant, zero exterior velocity, reflection and periodic conditions \\
      Input & VTK-family (.vtk .vti .vtr .vts .vtu .vtp .vtm .vtkhdf .pvd), Tecplot (.dat .plt), Nek5000/NekRS case (.nek5000 .nek .fxxxxx), Ensight Gold (.case), OpenFOAM case (.foam) \\
      Output & VTK scalar fields and metadata \\
      Execution & Multicore CPU and device-resident CUDA paths with MPI support \\
      Platform & Windows with a graphical interface, Linux command-line execution, multi-node Linux HPC systems \\
    \end{tabular}
    \caption{Capabilities of the \ftlesoftware{} and \fslesoftware{} modules.}
    \label{tab:software}
  \end{center}
\end{table}

The modules provide explicit Euler, second-order, fourth-order, and sixth-order Runge--Kutta integration, denoted as Euler, RK2, RK4 and RK6. The seven-stage RK6 follows schemes of \citet{Butcher1964}, with coefficients given in Appendix~\ref{app:rk6}. The nominal trajectory step is specified as a fraction of the interval between velocity snapshots. On regular velocity lattices, spatial interpolation can be bilinear/trilinear, cubic Hermite, tricubic \citep{LekienMarsden2005} reconstruction. Besides, a fifth-order weighted essentially non-oscillatory (WENO)-like reconstruction is also implemented referring to \citet{JiangShu1996} and \citet{Shu2009} for shock capture, avoiding smoothing and shadows in Lagrangian results.

Native unstructured-mesh variants use the interpolation weights of the containing cell. High-order stencils obtain any samples outside the domain through the selected boundary continuation. Velocity is linearly interpolated between adjacent snapshots at each Runge--Kutta stage.

Four boundary continuations are available. Constant extension samples the nearest boundary value outside the domain. Homogeneous exterior treatment returns zero velocity beyond the boundary. Reflection mirrors spatial indices at a face. Periodic treatment wraps the sampling stencil, and the FSLE distance calculation additionally uses the periodic minimum image.

FTLE advection begins from a regular diagnostic lattice and produces forward and backward flow maps. The solver computes the tensor in equation~\eqref{eq:cg}. A fixed temporal interval produces one dataset, which is also known as steady FTLE. Overlapping temporal windows produce a time-indexed collection of FTLE fields, which can inherit the time attribute of the flow field, and is sometimes referred to as unsteady FTLE.

Advection of FSLE begins from an anchor time with a central lattice and one displaced copy per neighbour direction. Central and neighbouring trajectories are advanced together until a crossing occurs, the trajectories become invalid, or the integration horizon is reached. Completed seeds are excluded from further event processing. Each directional result further contains the exponent, signed separation time and a status that distinguishes a successful crossing, a censored sample and invalid or escaped trajectories.

\ftlesoftware{} and \fslesoftware{} can either run in \textsc{StreamCenter+} as GUI module, or in a standalone command-line (CLI) mode.
Results are written in standard VTK format for visualisation in \textsc{StreamCenter+} and other compatible flow visualization tools like \texttt{Paraview}.

\subsection{High-throughput execution strategy}
\label{sec:optimisation}

Modern flow dataset can often come with large mesh count. Eulerian diagnostic methods often require fewer resources because they do not involve cross-snapshot actions such as time integration. However, for Lagrangian diagnostics, a high-throughput design is essential to enable large-scale analyses with tens of billions of degrees of freedom.
Parallel computing, data reuse, memory structure optimization, and other techniques have significantly reduced the overhead of such computations. At the same time, thanks to the development of GPUs, GPU acceleration has also become a powerful tool for scientific computing. The device-resident strategy can minimize the information exchange between the host and device, thereby improving computational efficiency. Profiling technology can identify potential optimization directions, making the memory structure and computational logic more reasonable, and further squeezing the efficiency headroom. 
Our implementation benefits from NVIDIA Nsight Systems as device profiler. The profiling details are not detailed here.

A direct calculation of many overlapping FTLE windows can treat every anchor time as an independent run. That approach repeatedly traverses the same velocity sequence and reconstructs identical temporal interpolation state. Instead, let $t_a$ denote an anchor and let $[t_k,t_{k+1}]$ be a source-data interval. During a forward sweep, only anchors in the active set
\begin{equation}
  \mathcal{A}_k^{+}=\left\{a:
  [t_k,t_{k+1}]\cap[t_a,t_a+T]\neq\varnothing\right\}
  \label{eq:active-ftle}
\end{equation}
are advanced through that interval. The backward sweep is defined analogously. Two adjacent velocity snapshots form a rolling temporal window and are shared by all active anchors in a batch. Thus a snapshot sequence is traversed monotonically rather than once per anchor. The number of simultaneous anchors is limited by an explicit memory budget with trimming, so the working set remains bounded when either the seed lattice or number of windows is extremely large, reducing the memory cost and avoiding out-of-memory (OOM) events.

Integration starts at the prescribed anchor time, including times between velocity snapshots. If a nominal step would pass $t_a+T$, the final step is shortened to end at that time. At every Runge--Kutta stage, velocity is linearly interpolated at the physical stage time. 

Particle coordinates are stored as separate contiguous component arrays. This layout supports unit-stride traversal, parallel first-touch placement and vector evaluation of the common interpolation path. A run-time instruction check selects an available vector kernel when the processor and operating system support it. Otherwise, the executable uses the scalar kernel. The default affinity policy distributes workers across physical cores and NUMA nodes before assigning simultaneous multithreading siblings, with bindings specified across processor groups. These measures are particularly important for high-order interpolation, for which arithmetic and stencil traffic dominate trajectory advancement.

FSLE offers an additional algorithmic opportunity because a seed has no further influence once its first valid crossing has been recorded. After each accepted step, completed, invalid and escaped seeds are removed from the active set. Neighbour trajectories are removed from the active set when they become invalid or cross an open boundary. Central and neighbour advancement, periodic distance evaluation and the event test are fused into one parallel traversal. If $\mathcal{S}_n$ is the unfinished seed set after step $n$, the dominant work is therefore proportional to
\begin{equation}
  W_{\mathrm{FSLE}}\sim
  \sum_n\sum_{i\in\mathcal{S}_n}
  \left(1+|\mathcal{N}_{i,n}|\right)s_{\mathrm{RK}},
  \label{eq:fsle-work}
\end{equation}
where $s_{\mathrm{RK}}$ is the number of Runge--Kutta stages and $\mathcal{N}_{i,n}$ is the still-relevant neighbour set. This replaces uniform advancement of every seed and every neighbour to the maximum horizon. It produces the largest benefit when many events occur early, while still avoiding unnecessary event processing in censored calculations.

\begin{figure}[t]
  \centering
  \includegraphics[width=0.70\textwidth]{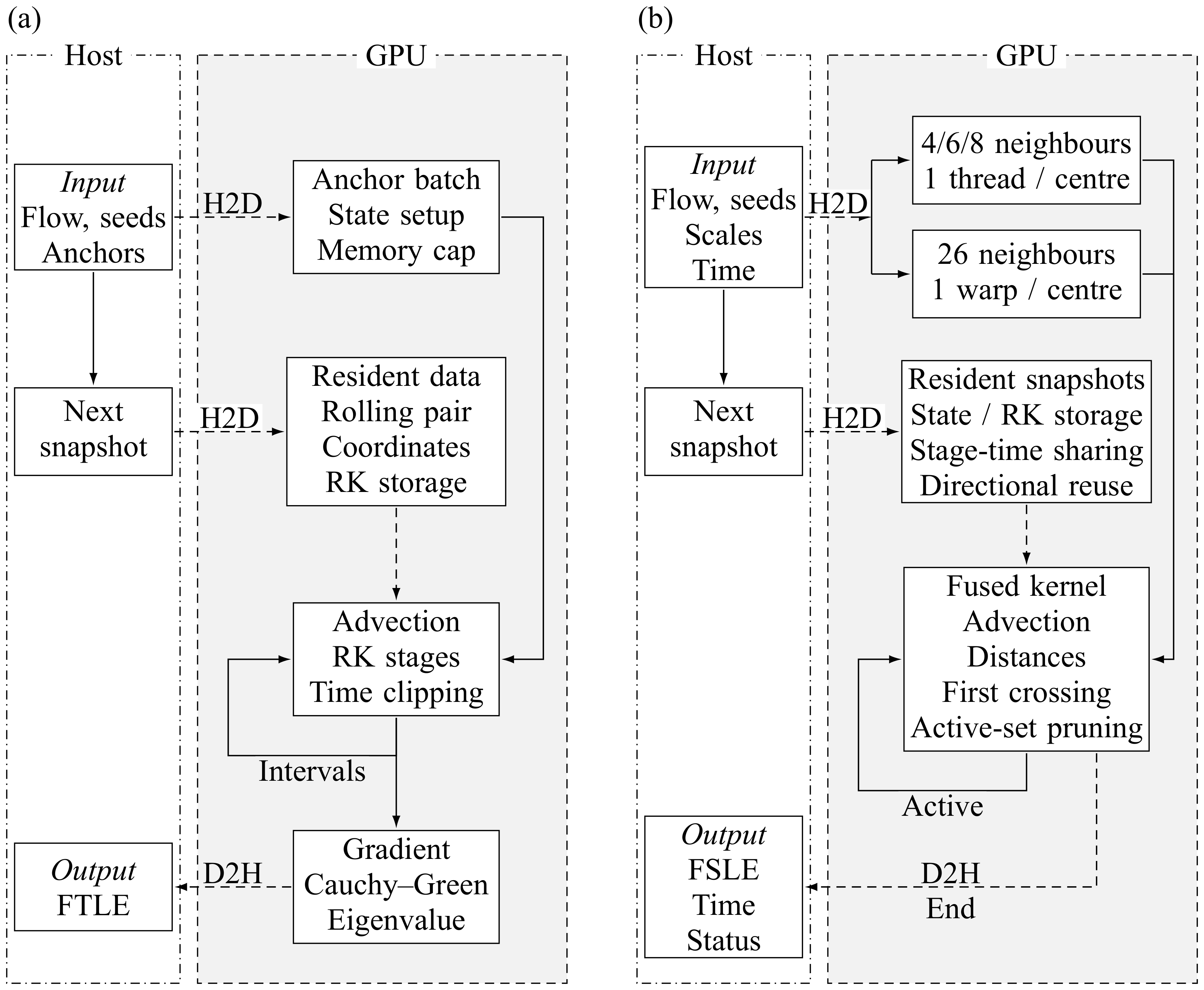}
  \caption{Device-resident GPU workflows for (a) FTLE and (b) FSLE. }
  \label{fig:gpu-workflows}
\end{figure}

Figure~\ref{fig:gpu-workflows} summarises the host--device data flow and GPU execution stages of both diagnostics. On the GPU, trajectory state and reusable stage storage remain resident throughout a directional sweep. Only the next velocity snapshot enters the rolling pair, avoiding repeated reconstruction of particle state between source-data intervals. After FTLE advection, flow-map differentiation, symmetric-tensor assembly and largest-eigenvalue evaluation are performed on the device, so only the completed diagnostic field must return to host memory. Independent anchor ranges can be divided between multiple devices without inter-device communication because each FTLE window has its own flow map.

The FSLE kernel is specialised by spatial dimension, boundary rule, integration scheme and fixed neighbourhood size. For the four-, six- and eight-neighbour stencils, each GPU thread advances one central seed and stores the corresponding fixed-size neighbour arrays. For the 26-neighbour 3-D stencil, one warp advances each central seed and its neighbours. Stage-time interpolation is shared across the neighbours and the execution block size is selected from occupancy constraints. Forward and backward branches reuse the same resident velocity sequence and allocated workspaces.

\FloatBarrier

\subsection{Platform performance}
\label{sec:performance}

Strong scaling was measured on Arrhenius HPC with a fixed $1000^3$ velocity grid and $10^9$ diagnostic seeds, using MPI benchmark drivers that call \ftlesoftware{} and \fslesoftware{} CLI. Trajectories in a synthetic steady velocity field were advanced in double precision with trilinear interpolation and 16 Euler steps of $\Delta t=0.02$. FSLE used six axial neighbours, $\delta_0=5\times10^{-4}$ and $\delta_f/\delta_0=\exp(0.012)$. CPU runs used one MPI rank per node and up to 256 OpenMP threads per rank, while GPU runs used one rank per NVIDIA GH200 GPU and four GPUs per node. Each rank held the velocity field and advanced a disjoint seed partition. FTLE exchanged one layer of flow-map values through MPI, with host staging on the GPU.

\begin{figure}[!b]
  \centering
  \includegraphics[width=\textwidth]{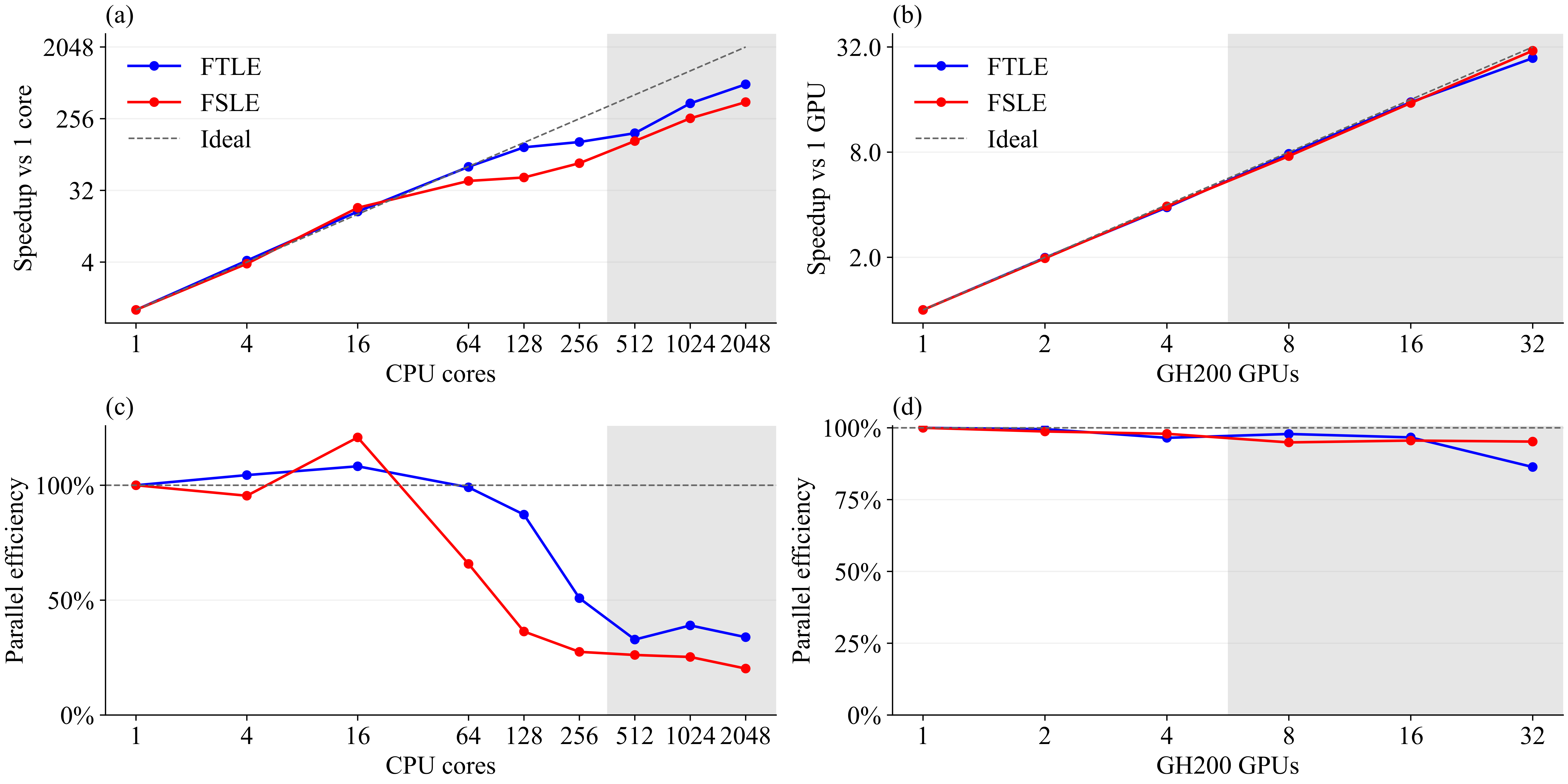}
  \caption{Strong scaling of the FTLE and FSLE numerical kernels for a $1000^3$ velocity grid and $10^9$ diagnostic seeds on Arrhenius. (a,b) Speed-up relative to one CPU core and one GPU, respectively. (c,d) Corresponding parallel efficiencies. Shading denotes multiple-node tests. Dashed lines indicate ideal scaling.}
  \label{fig:arrhenius-scaling}
\end{figure}

All timings are medians of three repetitions. The Arrhenius timings measure resident-data computation, including MPI communication, GPU synchronisation and the CPU FSLE routine's internal state initialisation. Input generation and file I/O are excluded. Figure~\ref{fig:arrhenius-scaling} reports speed-up $S(p)=t(1)/t(p)$ and parallel efficiency $E(p)=S(p)/p$, with separate single-core and single-GPU baselines.

At 2048 CPU cores, FTLE and FSLE required 1.747 and 13.830~s, respectively, giving speed-ups of 694 and 414 and efficiencies of 33.9\% and 20.2\%. GPU scaling remained nearly linear through 16 devices, with efficiencies of 96.7\% and 95.6\%. At 32 GPUs across eight nodes, compute times fell to 0.0298 and 0.0896~s, corresponding to satisfactory speed-ups of 27.63 and 30.47 and efficiencies of 86.3\% and 95.2\%. Thus both numerical kernels sustain efficient GPU scaling for a billion-seed workload, whereas CPU efficiency decreases more strongly with increasing concurrency, especially for multicore configurations but better for multiple-node setups.

Small-grid tests isolate the gains from the execution strategies. On an 80-core AMD EPYC 9V74 workstation with NVIDIA RTX 2080~Ti GPUs, the FSLE comparison used five ABC-flow snapshots on a $100^3$ velocity grid, $96^3$ seeds, periodic boundaries, trilinear interpolation and eight second-order Runge--Kutta steps over $T=0.4$, with 26 neighbours per seed. All seeds crossed for $\delta_f/\delta_0=1.01$, whereas none crossed for $\delta_f/\delta_0=4$. Table~\ref{tab:fsle-speed} shows CPU speed-ups of 17.93 and 3.91 for these cases, demonstrating the additional gain from removing completed seeds. A separate FTLE integration test reduced compute time from 0.329 to 0.0784~s with AVX2 command group, a factor of 4.20.

\begin{table}
  \begin{center}
    \begin{tabular}{llrrr}
      Branch and outcome & Backend & Reference (s) & Optimised (s) & Speed-up \\[3pt]
      Forward, early event & CPU  & 2.51 & 0.14 & 17.93 \\
      Forward, early event & GPU  & 0.23 & 0.12 & 1.92 \\
      Forward, censored    & CPU  & 2.50 & 0.64 & 3.91 \\
      Forward, censored    & GPU  & 0.59 & 0.48 & 1.23 \\
      Double-sided, early event  & GPU  & 0.31 & 0.20 & 1.55 \\
      Double-sided, censored     & GPU  & 1.03 & 0.95 & 1.08 \\
    \end{tabular}
    \caption{Compute-stage wall-times and speed-ups for the reference and optimised FSLE implementations.}
    \label{tab:fsle-speed}
  \end{center}
\end{table}

We also benchmark computational time and memory use against peer software packages (Table~\ref{tab:benchmark-software}). The tests use synthetic velocity fields on $256^2$ (for 2-D) and $256^3$ (for 3-D) velocity and seed grids, with 16 AMD EPYC 9755 CPU cores or four NVIDIA GH200 GPUs. SCP uses double precision with bilinear or trilinear interpolation. Fixed-step schemes use ten steps of $\Delta t=0.02$, while adaptive schemes retain their native step control. FSLE uses $\delta_0=5\times10^{-4}$ and $\delta_f/\delta_0=\exp(0.0075)$. Timings are medians of three runs after warm-up and cover numerical computation, device synchronisation and MPI communication, excluding input generation, file I/O and compilation.

\begin{table}
  \begin{center}
    \renewcommand{\arraystretch}{1.12}
    \begin{tabular}{@{}>{\raggedright\arraybackslash}p{0.335\textwidth}>{\centering\arraybackslash}p{0.14\textwidth}>{\centering\arraybackslash}p{0.16\textwidth}>{\centering\arraybackslash}p{0.115\textwidth}>{\raggedright\arraybackslash}p{0.18\textwidth}@{}}
      \textbf{Software} & \textbf{Computed\newline quantity} & \textbf{Computational\newline capability} & \textbf{Multi-node} & \textbf{Version} \\[3pt]
      \addlinespace[3pt]
      This work & FTLE \& FSLE & CPU \& GPU & Yes & 1.3.0 \\[3pt]
      \texttt{Dabiri LCS MATLAB Kit} \citep{DabiriLCSKit} & FTLE & CPU & No & 2.3 (BolltUpdate) \\[3pt]
      \texttt{LCStool} \citep{Onu2015} & FTLE & CPU & Yes & 1.0 \\[3pt]
      \texttt{TBarrier} \citep{TBarrier2023} & FTLE & CPU & No & 1.0.0 \\[3pt]
      \texttt{NumbaCS} \citep{JarvisRoss2025} & FTLE & CPU & No & 0.2.0 \\[3pt]
      \texttt{AVISO / lagrangian} \citep{AVISOLagrangian} & FTLE \& FSLE & CPU & Yes & 2.3.0 \\[3pt]
      \texttt{VisIt} \citep{Childs2012VisIt} & FTLE \& FSLE & CPU & Yes & 3.3.3 \\[3pt]
      \texttt{UVaFlow + UVaFTLE} \citep{UVaFlowRepository,UVaFTLE2023} & FTLE & CPU \& Mixed & No & UVaFlow: branch \mbox{2023-03-20};\newline UVaFTLE: branch \mbox{2025-01-28} \\[3pt]
      \texttt{Aquila-LCS} \citep{Lagares2024} & FTLE \& FSLE & CPU \& GPU & Yes & v1.0 \\
    \end{tabular}
    \caption{Software packages used in the benchmarks.}
    \label{tab:benchmark-software}
  \end{center}
\end{table}

\begin{figure}[t]
  \centering
  \includegraphics[width=\textwidth]{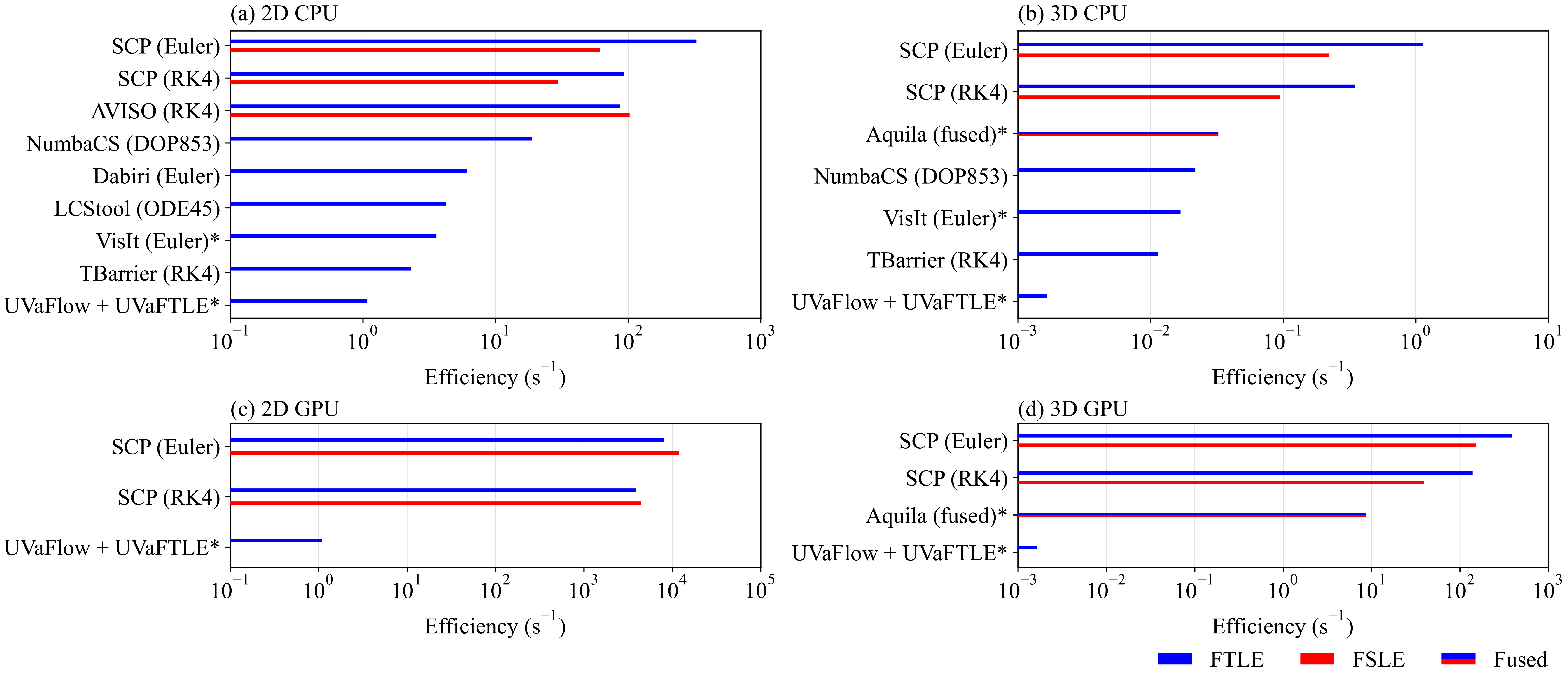}
  \caption{Computational efficiency $1/t$, where $t$ is the median compute time in seconds. (a,b) 2-D and 3-D tests with 16 CPU cores. (c,d) Corresponding tests with four GPUs. Blue and red denote FTLE and FSLE. The split-colour bars denote a single fused \texttt{Aquila-LCS} computation. Asterisks indicate failed numerical correctness checks for \texttt{Aquila-LCS}, \texttt{UVaFlow}--\texttt{UVaFTLE}, and \texttt{VisIt}.}
  \label{fig:solver-runtime-comparison}
\end{figure}

Figure~\ref{fig:solver-runtime-comparison} shows that SCP achieves the shortest measured 3-D runtimes on both CPU and GPU with either Euler or RK4 integration. For 3-D CPU FTLE, SCP with RK4 takes 2.87~s, compared with 46.05~s for \texttt{NumbaCS} and 87.50~s for \texttt{TBarrier}. On four GPUs, SCP computes 3-D FTLE and FSLE in 2.60 and 6.59~ms with Euler, and 7.23 and 25.81~ms with RK4.

The comparison retains each package's native diagnostic and integration method. \texttt{AVISO/lagrangian} evaluates a Cauchy--Green stretching exponent at the first separation event \citep{AVISOLagrangian}, whereas \texttt{Aquila-LCS} combines FTLE with a fixed-time mean neighbour-distance growth diagnostic \citep{AquilaRepository}. These differ from the first-separation-time FSLE in (\ref{eq:fsle}). \texttt{TBarrier} uses auxiliary trajectories with RK4, while \texttt{NumbaCS} uses adaptive DOP853 integration \citep{TBarrier2023,JarvisRoss2025}. The \texttt{UVaFlow}--\texttt{UVaFTLE} timings include CPU trajectory integration, followed by CPU or GPU diagnostic evaluation \citep{UVaFTLE2023}. 
It is particularly noteworthy that their FTLE diagnostic itself has very high computational efficiency, but the separated design of advection and diagnostic evaluation significantly slows down the overall computational efficiency of this workflow.

\begin{figure}[t]
  \centering
  \includegraphics[width=\textwidth]{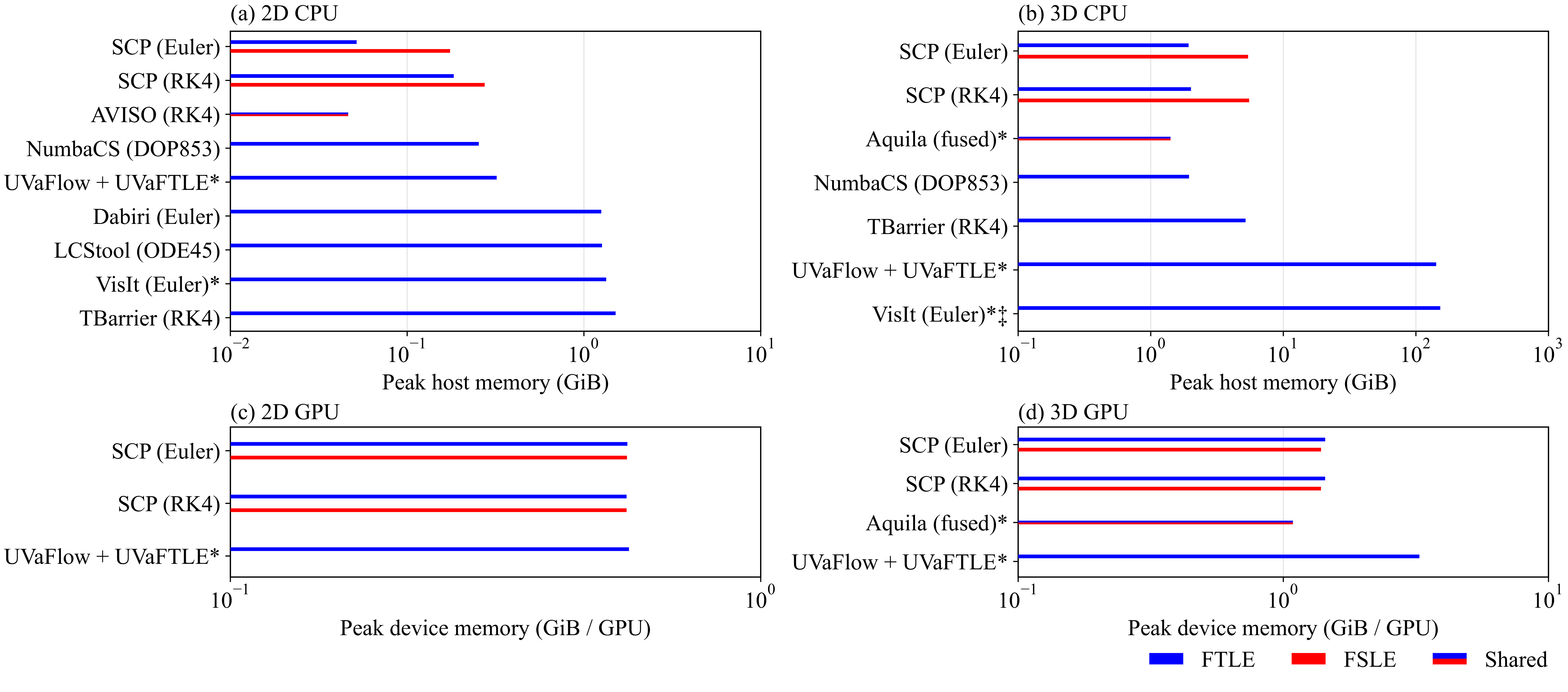}
  \caption{Benchmark of memory use. CPU panels show total job memory, including file cache. GPU panels show the largest single-device peak across four GPUs, including CUDA contexts. Values are sampled at 50~ms intervals, except for \texttt{VisIt} in 3-D ($\ddagger$), which uses the kernel-recorded peak. Split-colour bars indicate a shared FTLE/FSLE measurement. }
  \label{fig:solver-memory-comparison}
\end{figure}

Figure~\ref{fig:solver-memory-comparison} further reports the corresponding memory footprints. In the 3-D Euler tests, SCP uses 1.93~GiB of host memory for FTLE and 5.43~GiB for FSLE, with maximum per-GPU footprints of 1.44 and 1.39~GiB, respectively. RK4 gives similar 3-D memory footprints, so the additional integration stages primarily increase arithmetic work rather than storage, benefiting from our memory-trimming implementation.

\FloatBarrier

\section{Numerical verification}
\label{sec:validation}

\subsection{FTLE verification in steady and unsteady ABC flows}
\label{sec:abc}

The Arnold--Beltrami--Childress (ABC) flow provides a 3-D test containing both regular trajectory regions and chaotic stretching \citep{Dombre1986}. We consider the steady and time-dependent configurations of \citet{Haller2001} and compare the computed FTLE fields with the direct maximal finite-time Lyapunov exponent plots in that study. On the triply periodic domain $[0,2\pi]^3$, the velocity is
\begin{equation}
\begin{split}
  u(x,y,z,t) &= A(t)\sin z+C\cos y,\\
  v(x,y,z,t) &= B\sin x+A(t)\cos z,\\
  w(x,y,z,t) &= C\sin y+B\cos x,
\end{split}
\label{eq:abc}
\end{equation}
where $B=\sqrt{2}$ and $C=1$. The two cases differ only in the amplitude
\begin{equation}
 A(t)=
 \begin{cases}
  \sqrt{3}, & \text{steady ABC flow},\\[2pt]
  \sqrt{3}+\tfrac{1}{2}t\sin(\pi t), & \text{unsteady ABC flow}.
 \end{cases}
 \label{eq:abc-amplitude}
\end{equation}
Both fields are divergence free. The growing oscillation of $A(t)$ introduces an explicit time dependence while retaining spatial periodicity. All coordinates and times used here are nondimensional.

The analytic velocity is sampled on a uniform $100^3$ Cartesian lattice and supplied directly to \ftlesoftware{}. The steady case uses two identical snapshots spanning $0\leq t\leq10$, whereas the unsteady case uses 81 snapshots at intervals of $0.1$ over $0\leq t\leq8$. In each case, trajectories start at $t_0=0$ on a uniform $192^3$ seed lattice, and are advanced forward in time. The integration horizons are $T=10$ and $T=8$ for the steady and unsteady cases, respectively. The calculations use double precision on the GPU, tricubic spatial interpolation, linear interpolation between velocity snapshots, and RK6 integration with a physical time step of $0.02$. Periodic continuation is applied in all three spatial directions.
After advection, second-order finite differences of the flow map give the deformation gradient.

\begin{figure}[t]
  \centering
  \includegraphics[width=\textwidth]{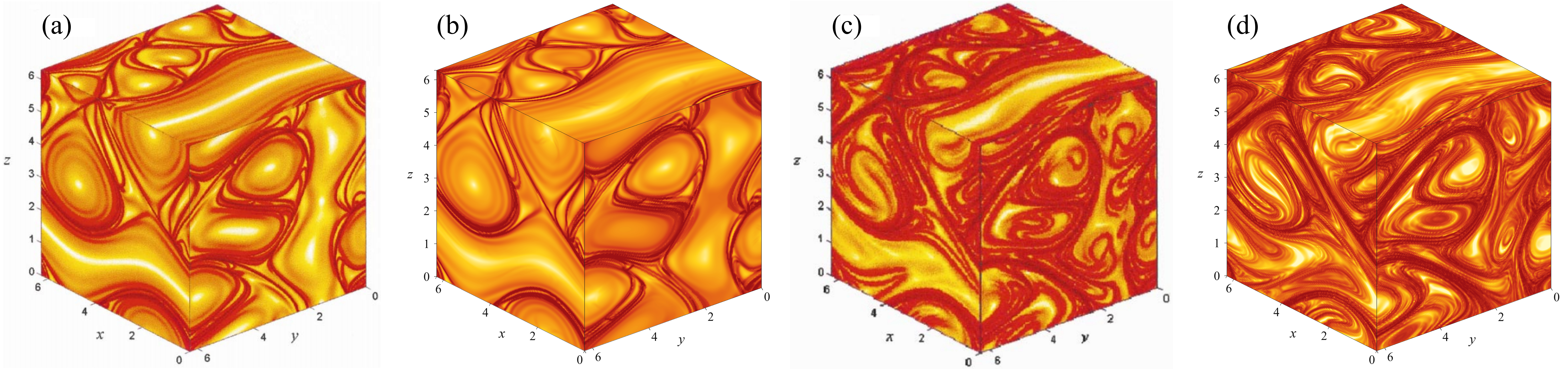}
  \caption{FTLE$_+$ in the ABC flows defined by (\ref{eq:abc})--(\ref{eq:abc-amplitude}). (a,b) Steady flow over $[0,10]$, with (a) \citet{Haller2001} and (b) the \ftlesoftware{} result. (c,d) Unsteady flow over $[0,8]$, with (c) \citet{Haller2001} and (d) the \ftlesoftware{} result. Darker colours indicate larger values.}
  \label{fig:abc-verification}
\end{figure}

Figure~\ref{fig:abc-verification}(a,b) compares \citet{Haller2001} and computed FTLE fields for the steady ABC flow. The computed field reproduces the broad low-stretching regions and the surrounding bands of elevated FTLE, including their continuation across the three visible faces. Figure~\ref{fig:abc-verification}(c,d) further shows the corresponding comparison for the unsteady flow. The time-dependent forcing produces a more intricate distribution of stretching, with the principal high-FTLE bands and intervening lower-FTLE regions following the spatial pattern by \citet{Haller2001}.

\FloatBarrier

\subsection{FTLE verification in stationary concentrated vortex model}
\label{sec:scvm}
 
The stationary concentrated vortex model (SCVM) provides a nonlinear 3-D test with radial inflow, axial transport and azimuthal motion \citep{Onishchenko2021}. We compare two configurations with \citet{Aksamit2023}. 

With $r=(x^2+y^2)^{1/2}$ and $E=\exp(-z/L-r^2/r_0^2)$, the cylindrical velocity components are
\begin{equation}
\begin{aligned}
u_r&=-v_0\frac{r}{L}\left(1-\frac{z}{L}\right)E,\qquad
u_\theta&=v_{\theta0}\frac{r}{r_0}\frac{z}{L}E,\qquad
u_z&=2v_0\frac{z}{L}\left(1-\frac{r^2}{r_0^2}\right)E.
\end{aligned}
\label{eq:scvm}
\end{equation}
Here $L=r_0=10$. The azimuthal component retains the factor $r/r_0$ in the original model, which makes the Cartesian velocity smooth at the axis. The analytic field is sampled on a Cartesian velocity lattice.

\begin{figure}[t]
  \centering
  \includegraphics[width=\textwidth]{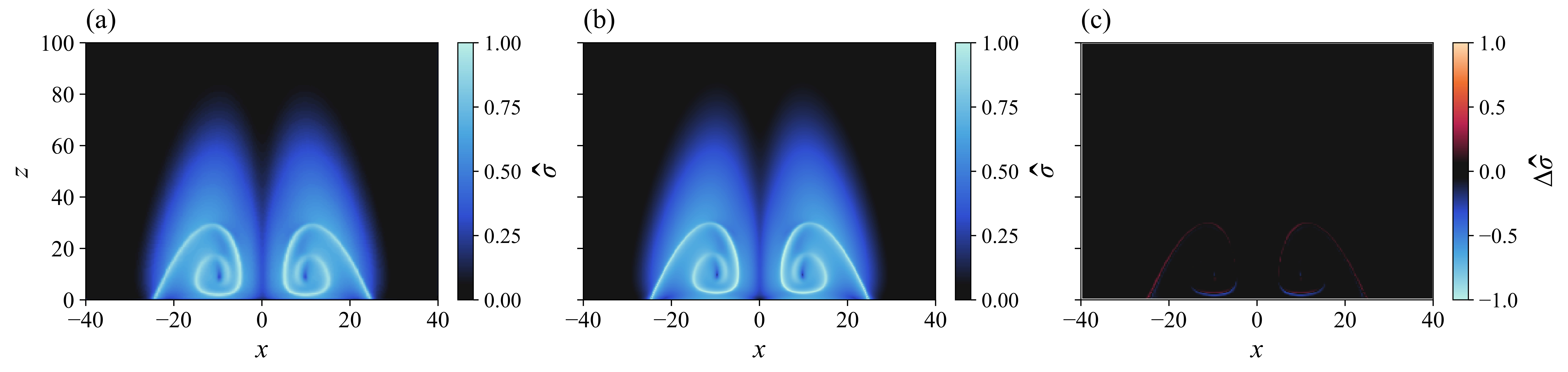}
  \caption{SCVM longitudinal section at $T=500$. (a) \citet{Aksamit2023}, reproduced under CC BY 4.0. (b) \ftlesoftware{}. (c) The normalised image difference, (b) minus (a).}
  \label{fig:scvm-section}
\end{figure}

For the longitudinal section, $v_0=1$, $v_{\theta0}=9$ and $T=500$. For Figure~\ref{fig:scvm-section}, \ftlesoftware{} evaluates the flow-map gradient using particles displaced by $10^{-3}$ from each seed. These perturbations are reset every 25 time units, and the local gradients are composed by the chain rule. RK4 integration with step $0.25$ and tricubic interpolation on a velocity lattice of spacing $0.25$ are used. The full 3-D gradient is evaluated on a $401\times501$ display grid in the $y=0$ plane. Against an independent cylindrical variational-equation reference integrated with DOP853, our FTLE result has a relative $L_2$ error of $0.0145\%$ over about 200,000 interior samples.

\begin{figure}[t]
  \centering
  \includegraphics[width=\textwidth]{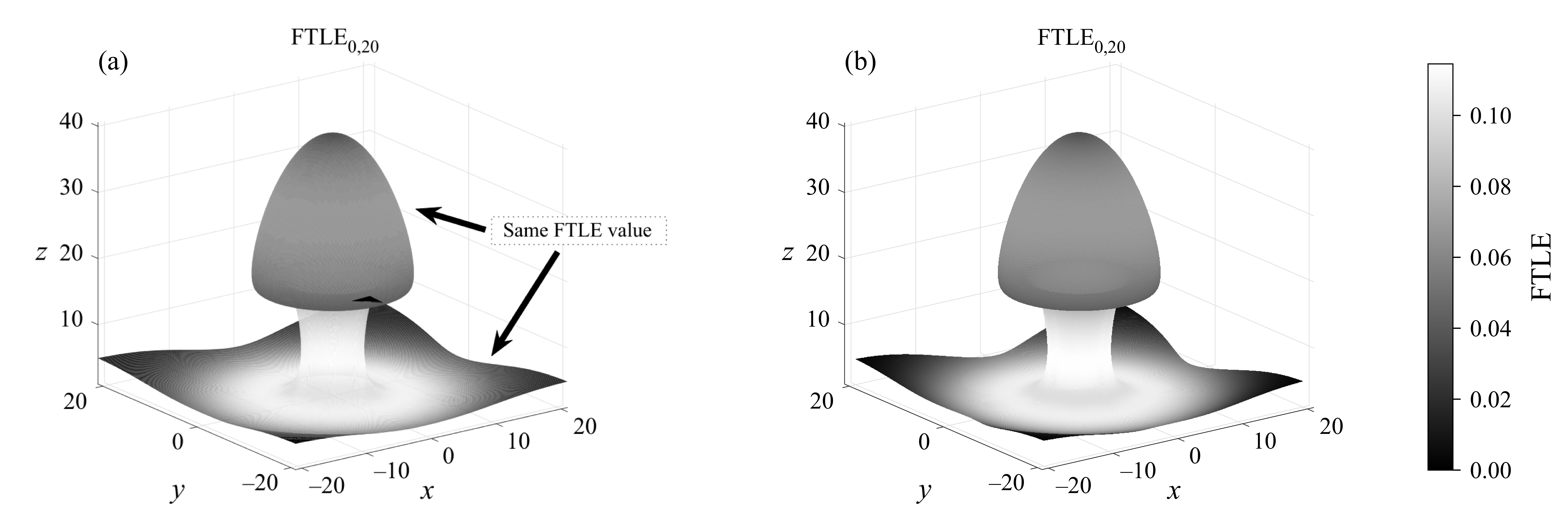}
  \caption{Advected $z_0=5$ SCVM material sheet at $T=20$: (a) \citet{Aksamit2023}, reproduced under CC BY 4.0. (b) \ftlesoftware{}.}
  \label{fig:scvm-sheet}
\end{figure}

A second configuration uses $v_0=5$, $v_{\theta0}=1$ and $T=20$. We advect a material sheet initially at $z_0=5$, using velocity spacing $0.5$, tricubic interpolation and RK4 integration with step $0.025$. Seed layers at $z_0=5\pm0.001$ provide the centred difference of the flow map with respect to the initial height, complementing the in-plane derivatives on the $z_0=5$ sheet. Figure~\ref{fig:scvm-sheet} displays the initial $[-20,20]^2$ subset of the computed $[-25,25]^2$ sheet, coloured by the FTLE at its initial positions, following \citet{Aksamit2023}. The maximum height is approximately $40.59$, and the rising core and expanded upper cap reproduce the main published geometry. At 2500 regular interior samples, the computed FTLE differs from the independent variational reference by only $0.0856\%$ in relative $L_2$ norm.

\subsection{FSLE verification in linear stretching and rotational shear flows}
\label{sec:fsle-analytic-verification}

Linear stretching and rotational shear flows, both steady, incompressible and 3-D, are used to verify \fslesoftware{} directly against analytical first-arrival times of particle pairs with finite initial separation. The linear stretching flow tests the identification of the fastest separating direction under axis permutation and integration in both time directions. The rotational shear flow tests 3-D pair separation, a spatially varying FSLE distribution and reachability.

\begin{samepage}
All quantities are dimensionless. The initial centre positions form the Cartesian product of $x_0,y_0,z_0\in\{-1,-0.9,\ldots,0.9,1\}$. Each centre has six neighbours, $\boldsymbol{q}_{j,0}=\boldsymbol{x}_0\pm\delta_0\boldsymbol{e}_k$, where $k=x,y,z$ and $\boldsymbol{e}_k$ is a unit basis vector. With $\delta_0=1$, $\delta_f=2$, $t_0=0$ and $T_{\max}=4$, the forward and backward first-arrival times and FSLE are defined using positive elapsed time $s$ as
\begin{equation}
\begin{aligned}
\tau^{\pm}(\boldsymbol{x}_0)&=\inf\!\left\{s>0:\max_{1\leq j\leq6}
\left\|\boldsymbol{X}(\boldsymbol{q}_{j,0},\pm s)-\boldsymbol{X}(\boldsymbol{x}_0,\pm s)\right\|_2\geq2\right\},\qquad
\lambda^{\pm}&=\frac{\ln2}{\tau^{\pm}}.
\end{aligned}
\label{eq:fsle-first-arrival}
\end{equation}
\end{samepage}
Distances use the 3-D Euclidean norm $\|\Delta\boldsymbol{X}\|_2=(\Delta x^2+\Delta y^2+\Delta z^2)^{1/2}$, and the first arrival is the earliest event among the six neighbours.

The analytical velocity fields are sampled on a regular grid over $[-16,16]^3$ with spacing $h_u=0.5$ and supplied to \fslesoftware{} as nine identical steady snapshots spanning $[-4,4]$. The calculations use double precision, RK4 integration and trilinear spatial interpolation, with time steps $\Delta t=0.01,0.005,0.0025$. Threshold-crossing times are obtained by linear interpolation of pair distances between consecutive time steps. Trilinear interpolation reproduces the selected linear and bilinear velocity functions exactly, and all computed trajectories remain within the velocity domain.

Table~\ref{tab:fsle-analytic-cases} lists the velocity fields and analytical FSLE values. In the linear stretching case $A$ and its axis permutation $A_{\mathrm{perm}}$, the fastest forward separation occurs along the $x$ and $y$ axes, respectively, and backward separation is fastest along the $z$ axis in both cases. The first-arrival times at all centres are
\begin{equation}
\tau_A^+=\tau_{A_{\mathrm{perm}}}^+=\frac{\ln2}{0.4},\qquad
\tau_A^-=\tau_{A_{\mathrm{perm}}}^-=\frac{\ln2}{0.5}.
\end{equation}

For the rotational shear case $B$, let $\rho=(x_0^2+y_0^2)^{1/2}$. The pair distances in both time directions satisfy
\begin{equation}
d_{\pm x}(s)=d_{\pm y}(s)=1,\qquad
d_{\pm z}^{\,2}(s)=1+4\rho^2\sin^2(s/2).
\end{equation}
Within the reachable region listed in Table~\ref{tab:fsle-analytic-cases}, the principal branch of the inverse sine gives
\begin{equation}
\tau_B^+=\tau_B^-=2\arcsin\!\left(\frac{\sqrt{3}}{2\rho}\right),\qquad
\lambda_B^+=\lambda_B^-=\lambda_{\mathrm{exact}}(\rho)
=\frac{\ln2}{2\arcsin\!\left(\sqrt{3}/(2\rho)\right)}.
\label{eq:fsle-B-exact}
\end{equation}
For $\rho<\sqrt{3}/2$, the target separation is never attained. The boundary $\rho=\sqrt{3}/2$ corresponds to a tangential event at $\tau=\pi$. No centres lie on this boundary in the present grid. There are 4200 analytically reachable and 5061 unreachable centres, and all reachable points satisfy $\tau<\pi<T_{\max}$.

\begin{table}
\centering
\small
\setlength{\tabcolsep}{5pt}
\begin{tabular}{@{}lcccc@{}}
\toprule
Case & Velocity field $\boldsymbol{u}$ & \shortstack{FSLE+} & \shortstack{FSLE-} & Reachability \\
\midrule
$A$ & $(0.4x,0.1y,-0.5z)$ & $0.4$ & $0.5$ & All centres \\
$A_{\mathrm{perm}}$ & $(0.1x,0.4y,-0.5z)$ & $0.4$ & $0.5$ & All centres \\
$B$ & $(-yz,xz,1)$ & $\lambda_{\mathrm{exact}}(\rho)$ & $\lambda_{\mathrm{exact}}(\rho)$ & $\rho>\sqrt{3}/2$ \\
\bottomrule
\end{tabular}
\caption{Velocity fields, analytical FSLE values and reachability for the three test cases.}
\label{tab:fsle-analytic-cases}
\end{table}

\begin{figure}[tbp]
\centering
\includegraphics[width=\linewidth]{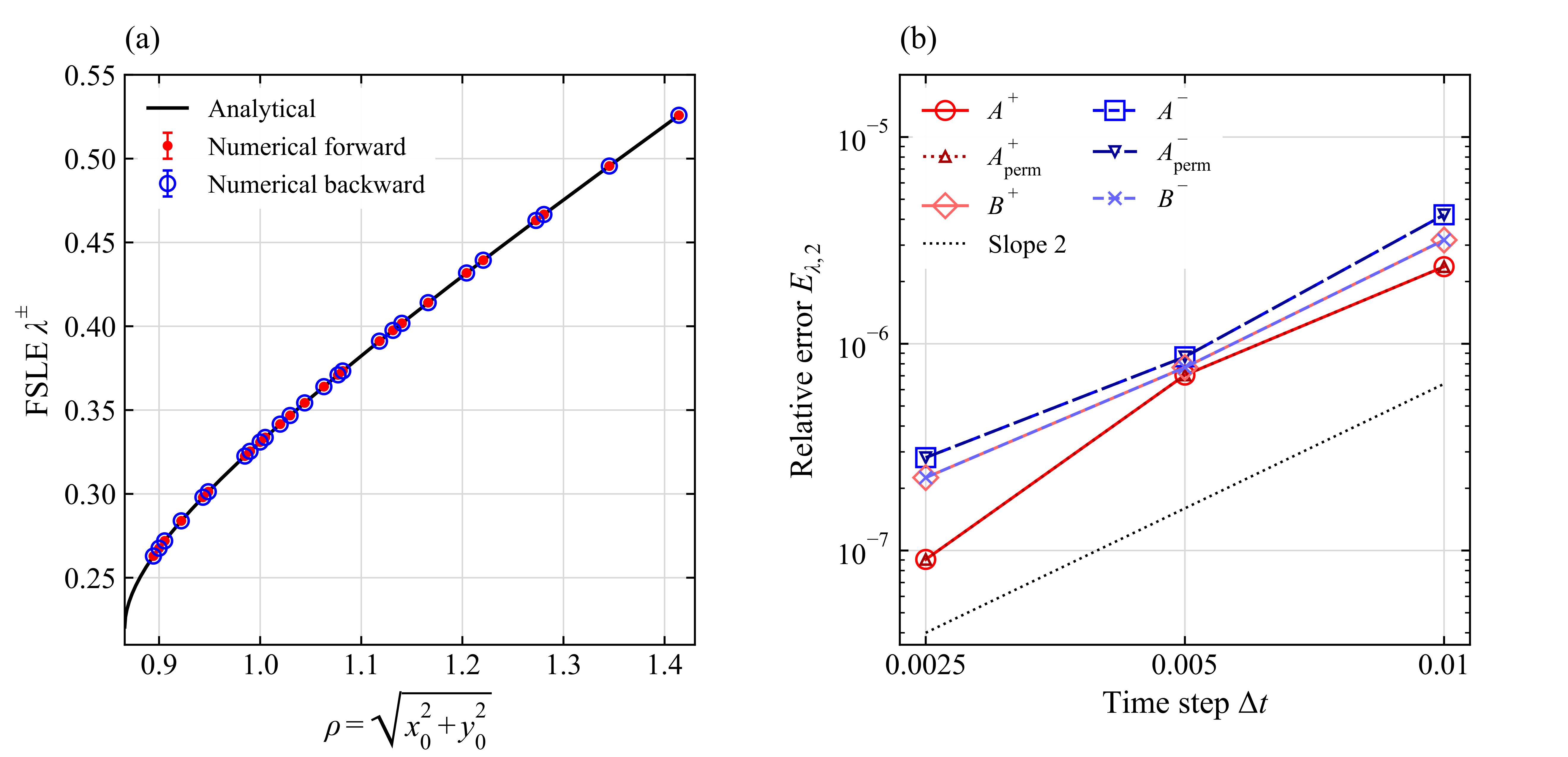}
\caption{Analytical verification of \fslesoftware{}. (a) Radial distribution for case $B$, and (b) time-step convergence.}
\label{fig:fsle-verification}
\end{figure}

Figure~\ref{fig:fsle-verification}(a) shows the radial distribution for case $B$ at the finest time step. Both forward and backward results agree with the analytical solution. Variations with axial position and azimuth at a fixed radius are small, reproducing the axial uniformity and rotational symmetry of the analytical FSLE.

\begin{samepage}
Over the analytically reachable set $R$, the relative errors in FSLE and first-arrival time are defined as
\begin{equation}
\begin{aligned}
E_{\lambda,2}&=\left[\frac{\sum_{i\in R}(\lambda_i^{\mathrm{num}}-\lambda_i^{\mathrm{exact}})^2}
{\sum_{i\in R}(\lambda_i^{\mathrm{exact}})^2}\right]^{1/2},\\
E_{\lambda,\infty}&=\max_{i\in R}\frac{|\lambda_i^{\mathrm{num}}-\lambda_i^{\mathrm{exact}}|}{|\lambda_i^{\mathrm{exact}}|},\qquad
E_{\tau,\infty}=\max_{i\in R}\frac{|\tau_i^{\mathrm{num}}-\tau_i^{\mathrm{exact}}|}{\tau_i^{\mathrm{exact}}}.
\end{aligned}
\end{equation}
\end{samepage}

All six error curves in Figure~\ref{fig:fsle-verification}(b) decrease as the time step is refined. The curves for $A$ and $A_{\mathrm{perm}}$ in the same time direction, and the forward and backward curves for $B$, overlap at the plotted scale. Although trajectories are integrated with a fourth-order scheme, linear interpolation of the first-arrival time still affects the overall FSLE accuracy. The observed order between the two finest levels is calculated as $p=\ln[E_{\lambda,2}(\Delta t)/E_{\lambda,2}(\Delta t/2)]/\ln2$. Differences between cases are associated with the location of the threshold crossing within a time step.

\begin{table}
\centering
\small
\setlength{\tabcolsep}{5pt}
\begin{tabular}{@{}lccccc@{}}
\toprule
Case/direction & $E_{\lambda,2}$ & $E_{\lambda,\infty}$ & $E_{\tau,\infty}$ & \shortstack{FP,FN} & Observed order $p$ \\
\midrule
$A/+$ & $9.05\times 10^{-8}$ & $9.05\times 10^{-8}$ & $9.05\times 10^{-8}$ & 0,0 & 2.96 \\
$A/-$ & $2.81\times 10^{-7}$ & $2.81\times 10^{-7}$ & $2.81\times 10^{-7}$ & 0,0 & 1.62 \\
$A_{\mathrm{perm}}/+$ & $9.05\times 10^{-8}$ & $9.05\times 10^{-8}$ & $9.05\times 10^{-8}$ & 0,0 & 2.96 \\
$A_{\mathrm{perm}}/-$ & $2.81\times 10^{-7}$ & $2.81\times 10^{-7}$ & $2.81\times 10^{-7}$ & 0,0 & 1.62 \\
$B/+$ & $2.25\times 10^{-7}$ & $5.64\times 10^{-7}$ & $5.64\times 10^{-7}$ & 0,0 & 1.77 \\
$B/-$ & $2.25\times 10^{-7}$ & $5.64\times 10^{-7}$ & $5.64\times 10^{-7}$ & 0,0 & 1.77 \\
\bottomrule
\end{tabular}
\caption{Relative FSLE and first-arrival-time errors at $\Delta t=0.0025$, reachability misclassifications, and convergence orders from the two finest time steps.}
\label{tab:fsle-finest-errors}
\end{table}

Table~\ref{tab:fsle-finest-errors} reports results for $\Delta t=0.0025$, with observed orders calculated from the $0.005$ and $0.0025$ levels. At the finest time step, the relative $L_2$ errors are below $3\times10^{-7}$, the maximum relative errors in FSLE and first-arrival time are below $6\times10^{-7}$. FP and FN denote the numbers of unreachable points classified as reachable and reachable points classified as unreachable, respectively. All centres are classified correctly as reachable or unreachable.

\FloatBarrier

\section{Examples of Application}
\label{sec:applications}

To demonstrate the applicability of our platform, two analysis cases are given here, spanning a large scale range. First one in \S\ref{sec:application-cavity} shows the development of 3-D coherent structures in a lid-driven cavity of infinite span, comparing a Newtonian fluid with a shear-thinning Carreau generalized Newtonian fluid by our pseudo-spectral solver. Then in \S\ref{sec:application-nus}, we present the FTLE results in a low-altitude wind field computed by our micrometeorological lattice Boltzmann method (LBM) code, taking National University of Singapore (NUS) campus as an example. It is worth noting that we are to demonstrate the applicability of our platform, therefore further deep dive into the flow mechanisms are not our main purpose in this section.

\subsection{Start-up non-Newtonian lid-driven cavity flow}
\label{sec:application-cavity}

3-D incompressible flow in a spanwise-periodic (infinite-length) lid-driven cavity is considered. Length, velocity, time and pressure are scaled by the cavity width $L$, the final lid-centre speed $U$, $L/U$ and $\rho U^2$, respectively. The dimensionless domain is $[0,1]\times[0,1]\times[0,0.8)$, with periodicity in $z$. With viscosity normalised by its dimensional zero-shear value $\mu_0$, the governing equations are
\begin{equation}
  \partial_i u_i=0,\qquad
  \partial_t u_i+u_j\partial_j u_i
  =-\partial_i p+\mathrm{Re}_0^{-1}\partial_j(2\mu S_{ij}).
  \label{eq:cavity-governing}
\end{equation}
Here $\partial_i=\partial/\partial x_i$, repeated indices are summed, $S_{ij}=(\partial_j u_i+\partial_i u_j)/2$, and $\mathrm{Re}_0=\rho UL/\mu_0=600$. The dimensionless Carreau viscosity is
\begin{equation}
  \mu=0.001+0.999[1+(10\dot{\gamma})^2]^{(n-1)/2},\qquad
  \dot{\gamma}=(2S_{ij}S_{ij})^{1/2}.
  \label{eq:cavity-viscosity}
\end{equation}
The Newtonian and shear-thinning cases use $n=1$ and $0.7$, respectively, with all other settings identical. Stationary no-slip conditions apply at $x=0,1$ and $y=0$. At the lid, $y=1$, the transverse velocities vanish and the streamwise velocity is
\begin{equation}
  u_{\mathrm{lid}}(x,t)
  =[1-(2x-1)^{16}]^2\{1-\cos[\pi\min(t,5)/5]\}/2.
  \label{eq:cavity-lid}
\end{equation}
Both cases start from the same divergence-free perturbation $\boldsymbol{u}(\boldsymbol{x},0)=\operatorname{curl}\boldsymbol{A}$, where
\begin{equation}
  \begin{gathered}
    \boldsymbol{A}=10^{-4}f(x)f(y)[0,\sin(2\kappa z),q(z)],\qquad
    f(s)=[1-(2s-1)^2]^2,\\[3pt]
    q(z)=\cos(\kappa z)+0.7\sin(2\kappa z)+0.3\cos(3\kappa z),\qquad
    \kappa=2\pi/0.8.
  \end{gathered}
  \label{eq:cavity-initial}
\end{equation}
The equations are solved using our in-house pseudo-spectral velocity--vorticity solver, \texttt{CoolBoxDNS}, with validation given in Appendix~\ref{app:coolbox-validation}. Chebyshev--Lobatto discretisation in $x$ and $y$ and a Fourier expansion in $z$ retain $65\times65\times33$ coefficients, respectively. Nonlinear and viscosity terms are evaluated on a padded $129\times129\times50$ grid, followed by an $L^2$ projection in the bounded directions and Fourier truncation. The grid was determined by a dependency check. Second-order Crank--Nicolson/Adams--Bashforth (CNAB) stepping treats diffusion with reference viscosity $\mu=1$ implicitly, and convection and the residual viscous stress explicitly. Both cases are advanced to $t=300$ with a fixed time step $\Delta t=0.001$.

\begin{figure}[t]
  \centering
  \includegraphics[width=\textwidth]{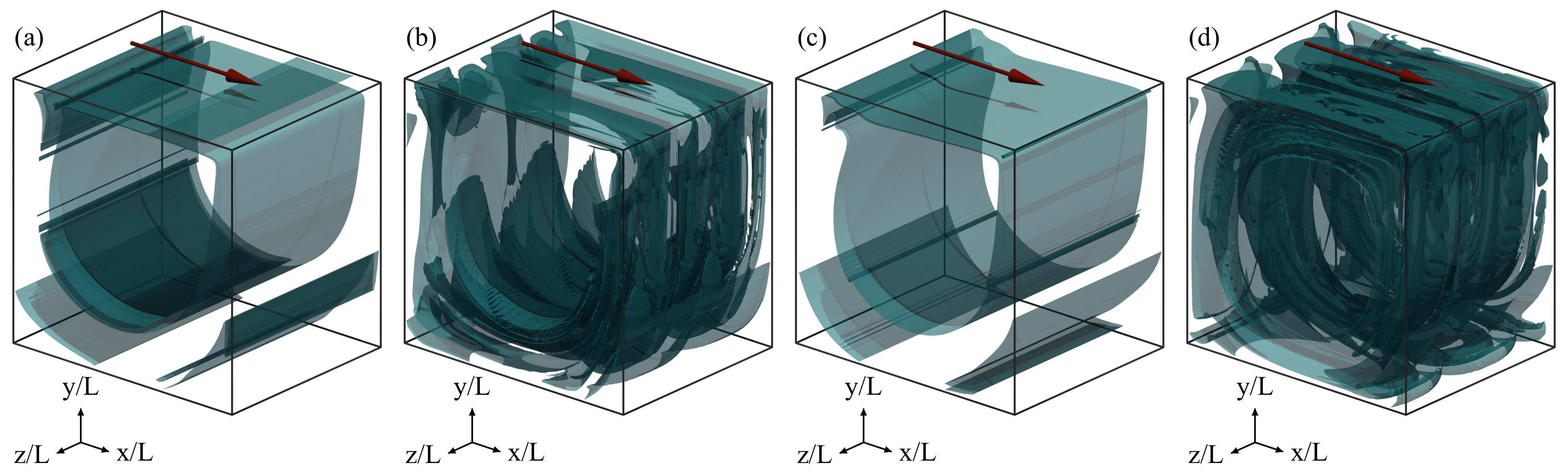}
  \caption{FTLE isosurfaces in the periodic lid-driven cavity at $t_0=280$ and $\sigma=0.23026\,U/L$. (a) Newtonian, $\mathrm{FTLE}_{+}$. (b) Carreau, $\mathrm{FTLE}_{+}$. (c) Newtonian, $\mathrm{FTLE}_{-}$. (d) Carreau, $\mathrm{FTLE}_{-}$.}
  \label{fig:cavity-isosurfaces}
\end{figure}

FTLE fields are computed with the structured-grid implementation of \ftlesoftware{}. Velocity snapshots spaced by $0.02$ are resampled to $257\times257\times100$ unique periodic velocity nodes using barycentric interpolation in $x$ and $y$ and Fourier interpolation in $z$. Trajectories use RK4 integration with a step of $0.005$, tricubic spatial interpolation and linear temporal interpolation. A $129\times129\times105$ seed grid is employed, including the duplicate periodic endpoint, with periodic particle wrapping and unwrapped flow-map coordinates in $z$.
Figure~\ref{fig:cavity-isosurfaces} presents the forward and backward fields at $t_0=280$ for both fluids. Corresponding planar sections are shown in Figures~\ref{fig:cavity-forward-sections} and \ref{fig:cavity-backward-sections} from Appendix~\ref{app:cavity-sections} for more FTLE details.

The identical initial perturbation develops differently in the two fluids. Figure~\ref{fig:cavity-modal-energy} from Appendix~\ref{app:cavity-sections} shows decay of the nonzero spanwise Fourier modes in the Newtonian flow, whereas the Carreau $m=2$ mode grows and higher harmonics develop. This contrast is consistent with the lower instability threshold of shear-thinning cavity flow reported by \citet{Haque2012}. At $t_0=280$, the Carreau flow has reached a finite-amplitude nonlinear regime.

The Newtonian FTLE isosurfaces form smooth sheets extended along $z$ and isolates a ``U''-shaped region in the upper cavity, which means the flows rotate isolately in this region, and the mass exchange with the surrounding fluid is suppressed.
While, the Carreau sheets become corrugated and develop wall-adjacent lobes projecting into the interior (Figure~\ref{fig:cavity-isosurfaces}). Figures~\ref{fig:cavity-forward-sections} and \ref{fig:cavity-backward-sections} from Appendix~\ref{app:cavity-sections} show that the central low-FTLE region persists, while the surrounding stretching layers vary along the span. The reorganisation is particularly pronounced in the backward FTLE: curved bands surround the central region in the $xy$ plane, and hook-shaped contours appear in the $xz$ and $yz$ planes.

\begin{figure}[!t]
  \centering
  \includegraphics[width=\textwidth]{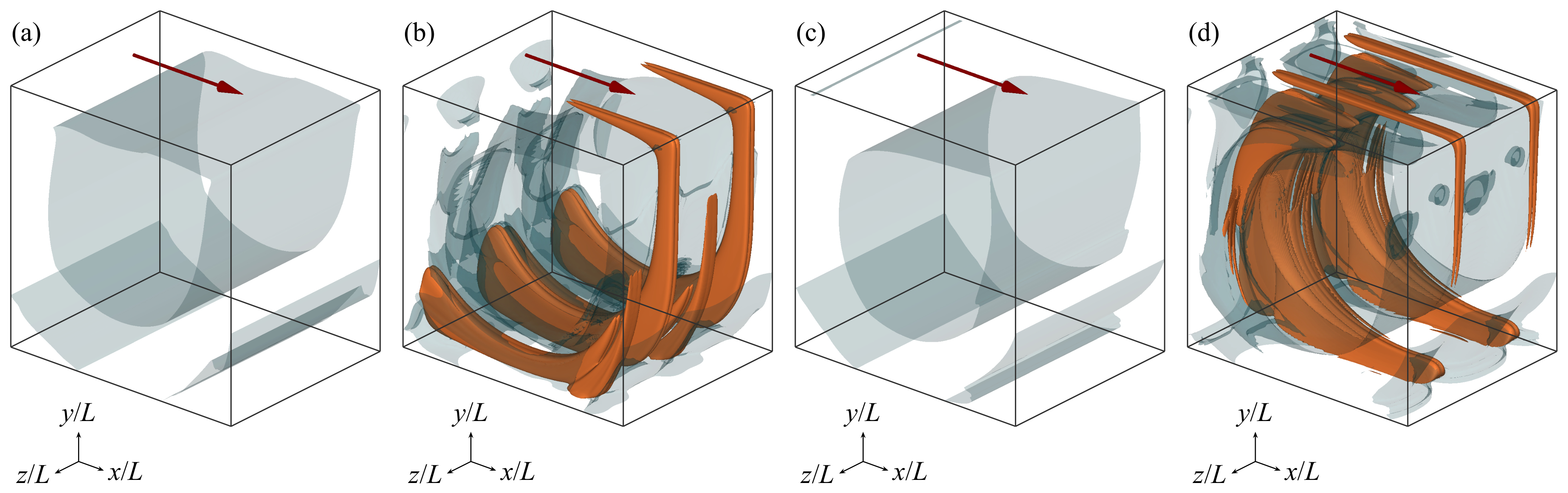}
  \caption{FSLE for first-arrival regions at $t_0=280$, with initial separation $\delta_0=0.02L$ and observation duration $H=5L/U$. (a,b) Forward FSLE analysis for Newtonian and Carreau flows; (c,d) corresponding backward analysis. Teal: fourfold growth of total separation using six axial neighbours. Orange: fourfold growth of spanwise separation using the two initially spanwise neighbours.}
  \label{fig:cavity-fsle-arrival}
\end{figure}

Figure~\ref{fig:cavity-velocity-sections} from Appendix~\ref{app:cavity-sections} shows the dominant cavity circulation in both fluids and secondary recirculation in the Carreau flow, repsented by velocity fields. The FTLE sections delineate narrow stretching layers and spanwise folds more distinctly than the instantaneous speed maps. FTLE relates the three-dimensional disturbance to accumulated material stretching \citep{Haller2011}. The FSLE analysis in Figure~\ref{fig:cavity-fsle-arrival} extends this comparison to the direction and finite scales of particle separation \citep{Aurell1997}. At $t_0=280$, the teal surfaces delimit initial positions from which total separation grows from $0.02L$ to $0.08L$ within $5L/U$. These total-separation FSLE first-arrival regions occur in both fluids. In the Newtonian flow, initially spanwise neighbours retain almost unchanged spanwise separation, consistent with the decay of the three-dimensional perturbation. Its substantial total separation therefore reflects deformation within the cavity cross-section, coexisting with negligible spanwise relative dispersion.

The spanwise-projected FSLE analysis additionally identifies localised orange first-arrival regions in the Carreau flow, where spanwise separation reaches four times its initial value. In the forward FSLE analysis, this target is reached over $5.2\%$ of the common sampled volume. Spanwise separation grows through differences in spanwise velocity experienced by neighbouring trajectories. Together with the modal-energy growth, these directional FSLE results connect the amplified disturbance to differential spanwise advection and finite-distance particle separation. The FTLE isosurfaces in Figure~\ref{fig:cavity-isosurfaces} reveal the corrugation of stretching layers. While the FSLE analysis in Figure~\ref{fig:cavity-fsle-arrival} additionally identifies where neighbouring particles separate by prescribed distances along the span. Backward spanwise-projected FSLE complements this understanding by identifying positions whose currently neighbouring particles had larger spanwise separations in the past, revealing their history of spanwise convergence. This directional, scale-dependent FSLE analysis distinguishes how the two fluids disperse neighbouring particles over finite distances, complementing the stretching geometry captured by FTLE.

The combined double-directional FSLE and FTLE analyses thus resolve stretching geometry and transport barriers, as well as spatially organised separation and convergence, which are associated with complex three-dimensional flows.

\FloatBarrier

\subsection{Low-altitude wind field at NUS campus}
\label{sec:application-nus}

This case considers low-altitude wind fields around the UTown at the National University of Singapore (NUS), obtained with our micrometeorological lattice Boltzmann method (LBM) solver \texttt{LatticeUrbanWind} \citep{LatticeUrbanWind2026}. Building footprints and terrain elevations define an observation area of approximately $1391\times1272$~m, with $x$ and $y$ pointing east and north. The computational domain extends 600~m westwards and 360~m northwards to provide velocity data for downstream trajectories. Building footprints \citep{3dglobfp} and terrain from a digital elevation model (DEM) \citep{NASADEM_2021} are voxelised into a geometric model represented by triangular surface elements. In this study, we adopt the D3Q19 discrete velocity model, where $D$ denotes the spatial dimension and $Q$ represents the number of discrete velocity directions. Large-eddy simulation (LES) was performed with Smagorinsky--Lilly model. The lattice resolution is 2~m. The solver mesh contains $996\times816\times273$ nodes, including a 90~m upper sponge layer. For more information on the numerical method and solver validation, please refer to our previous work \citep{LatticeUrbanWind2026,U3DWind2026}. 

The inlet uses the annual-mean vertical wind-speed profile from Singapore's 8760 hourly records for 2025 from ERA5 reanalysis \citep{Olauson2018ERA5}, with a prescribed meteorological wind direction of $120^\circ$. Superimposed von K\'arm\'an synthetic turbulence has an intensity of 7\% and a length scale of 70~m. The flow is advanced with $\Delta t\approx0.033$~s. After 1808~s start-up, three sources of radius 7~m release passive tracers 3--5~m above the local ground for 1200~s, at 128 particles per source every 2~s. Tracers use RK4 integration at the CFD time step.

\begin{figure}[!t] 
  \centering
  \includegraphics[width=\textwidth]{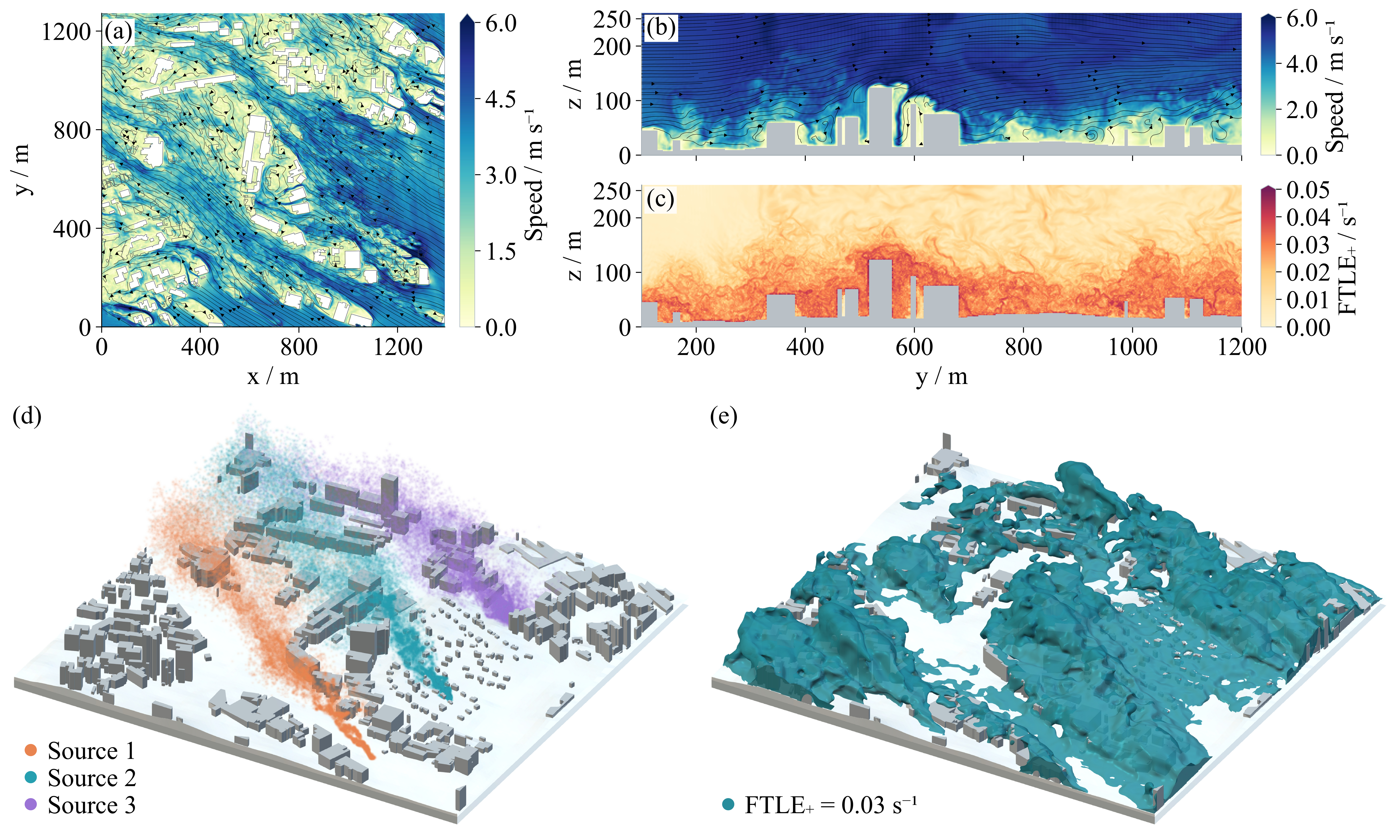}
  \caption{Wind and transport around NUS University Town at $t_0\approx3000$~s. (a) Speed and projected streamlines 30~m above local ground. (b,c) Speed and FTLE$_+$ in the vertical section. (d) Tracers from three sources. (e) FTLE$_+$ isosurface at the 95th percentile. }
  \label{fig:nus-wind-field}
\end{figure}

Using \ftlesoftware{}, forward FTLE is evaluated at $t_0\approx3000$~s over $T=120$~s on a $696\times636\times228$ observation grid, RK4 integration with a maximum step of 0.25~s, and no-slip-constrained piecewise multilinear spatial interpolation. Velocity snapshots are spaced by 1~s. Near walls, the selected treatment combines local step reduction with fluid-sided flow-map differences and stationary-wall constraints. Figure~\ref{fig:nus-wind-field} brings together the instantaneous velocity, tracer distribution and forward FTLE.

The velocity maps show wind channelled between buildings and reduced speeds in sheltered regions (Figure~\ref{fig:nus-wind-field}(a,b)). The three tracer clouds follow distinct downwind paths through the urban geometry (Figure~\ref{fig:nus-wind-field}(d)), simulating the transport of pollutants in the urban environment. The FTLE section reveals filamentary regions of enhanced stretching within and above the building canopy, while the isosurface resolves their 3-D organisation across the district (Figure~\ref{fig:nus-wind-field}(c,e)). 
The FTLE field indicates the presence of strong flow separation and mixing in densely populated areas, encompassing both the flow around buildings and the top regions. Conversely, in areas with fewer buildings like corridors, the FTLE structure is weaker. This can also be observed in particle tracing. Combining Figure~\ref{fig:nus-wind-field}(d) and (e), among the three sources we released, source 1 is located in a corridor, and its advection trajectory passes through fewer coherent structures, whereas source 2 and source 3 pass through more complex flow-around regions. Therefore, source 1 exhibits less particle diffusion, while source 2 and source 3 demonstrate earlier onset and stronger particle diffusion phenomena especially in the early stages.

\section{Conclusions}
\label{sec:conclusions}

This work has presented \ftlesoftware{} and \fslesoftware{} as an integrated platform for evaluating finite-time deformation and finite-scale particle separation from time-resolved velocity data. Support for structured and native unstructured meshes, selectable numerical schemes, and desktop and command-line execution connects these diagnostics to CFD workflows on workstations and multi-node systems. The principal computational contribution is the coordinated treatment of trajectory integration, data access and diagnostic evaluation: overlapping FTLE windows share rolling velocity snapshots within a bounded memory budget, GPU trajectories and intermediate fields remain device-resident, and FSLE active sets shrink as first-separation events are resolved.

The benchmarks show that these strategies support efficient volumetric computation. For a billion diagnostic seeds, FTLE and FSLE attain strong-scaling efficiencies of about 86\% and 95\%, respectively, on 32 GPUs. Multi-node CPU execution achieves speed-ups of several hundred relative to a single core. In the cross-package tests, SCP records the shortest measured 3-D compute times on both CPU and GPU with Euler and RK4 integration. The small-grid tests further demonstrate the benefits of CPU vectorisation for FTLE and of removing completed seeds from FSLE integration.

Numerical verification provides complementary checks of spatial structure, exponent accuracy and event detection. The ABC cases reproduce the published FTLE patterns in steady and unsteady flows. For both SCVM configurations, the computed FTLE agrees closely with independent variational-equation references. For FSLE, the linear stretching and rotational shear cases recover the analytical forward and backward first-arrival behaviour and correctly classify reachability at every tested centre.

The applications demonstrate how the computed fields expose accumulated deformation in complex 3-D flows. In the periodic cavity, the nearly spanwise-uniform stretching sheets of the Newtonian case become corrugated in the shear-thinning case, accompanying the growth of a finite-amplitude spanwise disturbance. In the urban wind field, FTLE resolves filamentary stretching regions within and above the building canopy, complementing the distinct dispersion patterns of the three tracer sources. Together, the two examples connect the development of three-dimensional flow structures with the spatial organisation of material stretching.

By combining verified diagnostics, numerical flexibility and scalable execution, the platform makes volumetric FTLE and FSLE analysis practical for large CFD datasets. Future work will focus on using LCS to track and attribute flow instability from the material transport perspective.

\section*{Funding}
This research recieved no funding.


\section*{Declaration of interests}
The authors report no conflict of interest.

\section*{Data availability statement}
Our software and examples are freely available at \url{https://github.com/hweifluids/StreamCenterPlus-Release} (will release upon acceptance of this paper).

\section*{Author ORCIDs}
H. Wei, \url{https://orcid.org/0000-0003-4286-8345};\\
M. Zhang, \url{https://orcid.org/0000-0002-8354-7129}.

\vfill

\clearpage
\appendix
\renewcommand{\theHsection}{\Alph{section}}
\section{Seven-stage sixth-order Runge--Kutta schemes}
\label{app:rk6}

The RK6(7) schemes use seven velocity evaluations per step and have order six \citep{Butcher1964}. For a signed time step $h$, the trajectory update is
\begin{equation}
\begin{aligned}
  \boldsymbol{k}_i&=\boldsymbol{u}\!\left(\boldsymbol{x}_n+h\sum_{j=1}^{i-1}a_{ij}\boldsymbol{k}_j,\,t_n+c_i h\right),\qquad i=1,\ldots,7,\\
  \boldsymbol{x}_{n+1}&=\boldsymbol{x}_n+h\sum_{i=1}^{7}b_i\boldsymbol{k}_i.
\end{aligned}
\label{eq:rk6-update}
\end{equation}
The coefficients used by \ftlesoftware{} are given below, with $q=\sqrt{5}$.
\begin{equation}
\renewcommand{\arraystretch}{1.8}
\setlength{\arraycolsep}{5pt}
\begin{array}{c|rrrrrrr}
c_i & a_{i1} & a_{i2} & a_{i3} & a_{i4} & a_{i5} & a_{i6} & a_{i7} \\ \hline
0 &  &  &  &  &  &  &  \\
\frac{5-q}{10} & \frac{5-q}{10} &  &  &  &  &  &  \\
\frac{5+q}{10} & \frac{-q}{10} & \frac{5+2q}{10} &  &  &  &  &  \\
\frac{5-q}{10} & \frac{-15+7q}{20} & \frac{-1+q}{4} & \frac{15-7q}{10} &  &  &  &  \\
\frac{5+q}{10} & \frac{5-q}{60} & 0 & \frac{1}{6} & \frac{15+7q}{60} &  &  &  \\
\frac{5-q}{10} & \frac{5+q}{60} & 0 & \frac{9-5q}{12} & \frac{1}{6} & \frac{-5+3q}{10} &  &  \\
1 & \frac{1}{6} & 0 & \frac{-55+25q}{12} & \frac{-25-7q}{12} & 5-2q & \frac{5+q}{2} &  \\
\hline
b_j & \frac{1}{12} & 0 & 0 & 0 & \frac{5}{12} & \frac{5}{12} & \frac{1}{12} \\
\end{array}
\label{eq:rk6-ftle-tableau}
\end{equation}

The coefficients used by \fslesoftware{} are
\begin{equation}
\renewcommand{\arraystretch}{1.8}
\setlength{\arraycolsep}{5pt}
\begin{array}{c|rrrrrrr}
c_i & a_{i1} & a_{i2} & a_{i3} & a_{i4} & a_{i5} & a_{i6} & a_{i7} \\ \hline
0 &  &  &  &  &  &  &  \\
\frac{1}{3} & \frac{1}{3} &  &  &  &  &  &  \\
\frac{2}{3} & 0 & \frac{2}{3} &  &  &  &  &  \\
\frac{1}{3} & \frac{1}{12} & \frac{1}{3} & \frac{-1}{12} &  &  &  &  \\
\frac{1}{2} & \frac{-1}{16} & \frac{9}{8} & \frac{-3}{16} & \frac{-3}{8} &  &  &  \\
\frac{1}{2} & 0 & \frac{9}{8} & \frac{-3}{8} & \frac{-3}{4} & \frac{1}{2} &  &  \\
1 & \frac{9}{44} & \frac{-9}{11} & \frac{63}{44} & \frac{18}{11} & 0 & \frac{-16}{11} &  \\
\hline
b_j & \frac{11}{120} & 0 & \frac{27}{40} & \frac{27}{40} & \frac{-4}{15} & \frac{-4}{15} & \frac{11}{120} \\
\end{array}
\label{eq:rk6-fsle-tableau}
\end{equation}

\clearpage
\section{Channel-flow validation of CoolBoxDNS}
\label{app:coolbox-validation}

\begin{figure}[htbp]
  \centering
  \includegraphics[width=\textwidth]{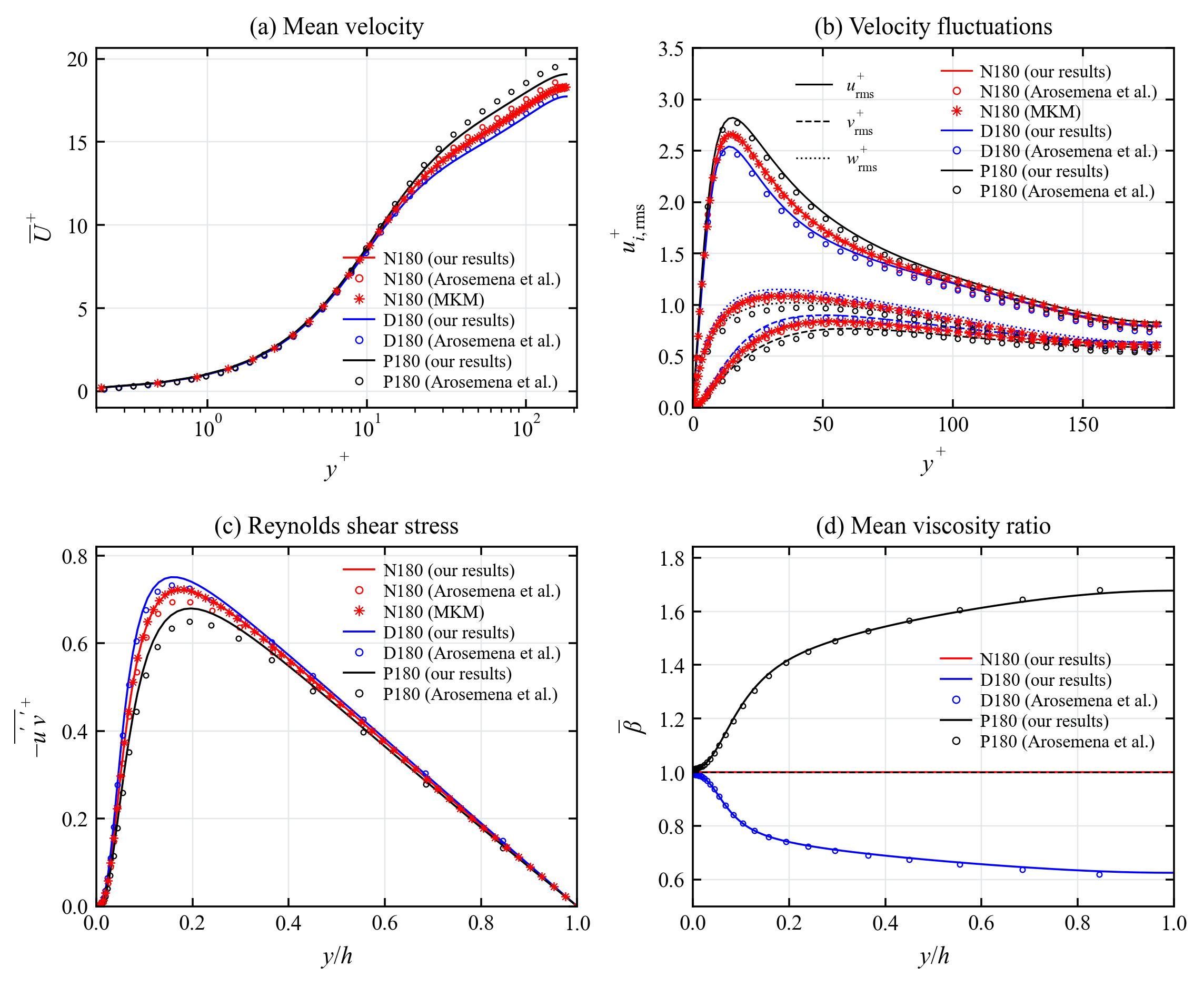}
  \caption{Channel-flow validation of \texttt{CoolBoxDNS}: (a) mean streamwise velocity, (b) root-mean-square velocity fluctuations, (c) Reynolds shear stress, and (d) mean viscosity ratio. Red, blue and black curves denote N180, D180 and P180, respectively. In (b), solid, dashed and dotted curves denote the streamwise, wall-normal and spanwise components. Open circles are digitized data from \citet{Arosemena2021}. Asterisks in (a--c) are the Newtonian reference data of \citet{moser1999turbchannel} (MKM).}
  \label{fig:coolbox-channel-validation}
\end{figure}

Figure~\ref{fig:coolbox-channel-validation} compares our \texttt{CoolBoxDNS} results with the turbulent channel-flow statistics of \citet{Arosemena2021} at $Re_\tau=u_\tau h/\nu_w\approx180$. The reference cases are Newtonian (N180), Carreau shear-thickening with $n=1.2$ (D180), and Carreau shear-thinning with $n=0.8$ (P180). The Newtonian case is also compared with the independent DNS data of \citet{moser1999turbchannel}. Here $h$ is the channel half-height, $y$ is the distance from the nearest wall, $u_\tau=\sqrt{\tau_w/\rho}$, and $\nu_w=\mu_w/\rho$. Superscript $+$ denotes wall units based on each case's mean wall stress $\tau_w$ and wall viscosity $\mu_w$, and $\overline{\beta}=\overline{\mu}/\mu_w$ is the mean viscosity ratio.

\clearpage

\section{Flow fields and disturbance energy in the periodic lid-driven cavity}
\label{app:cavity-sections}

\begin{figure}[htbp]
  \centering
  \includegraphics[width=\textwidth]{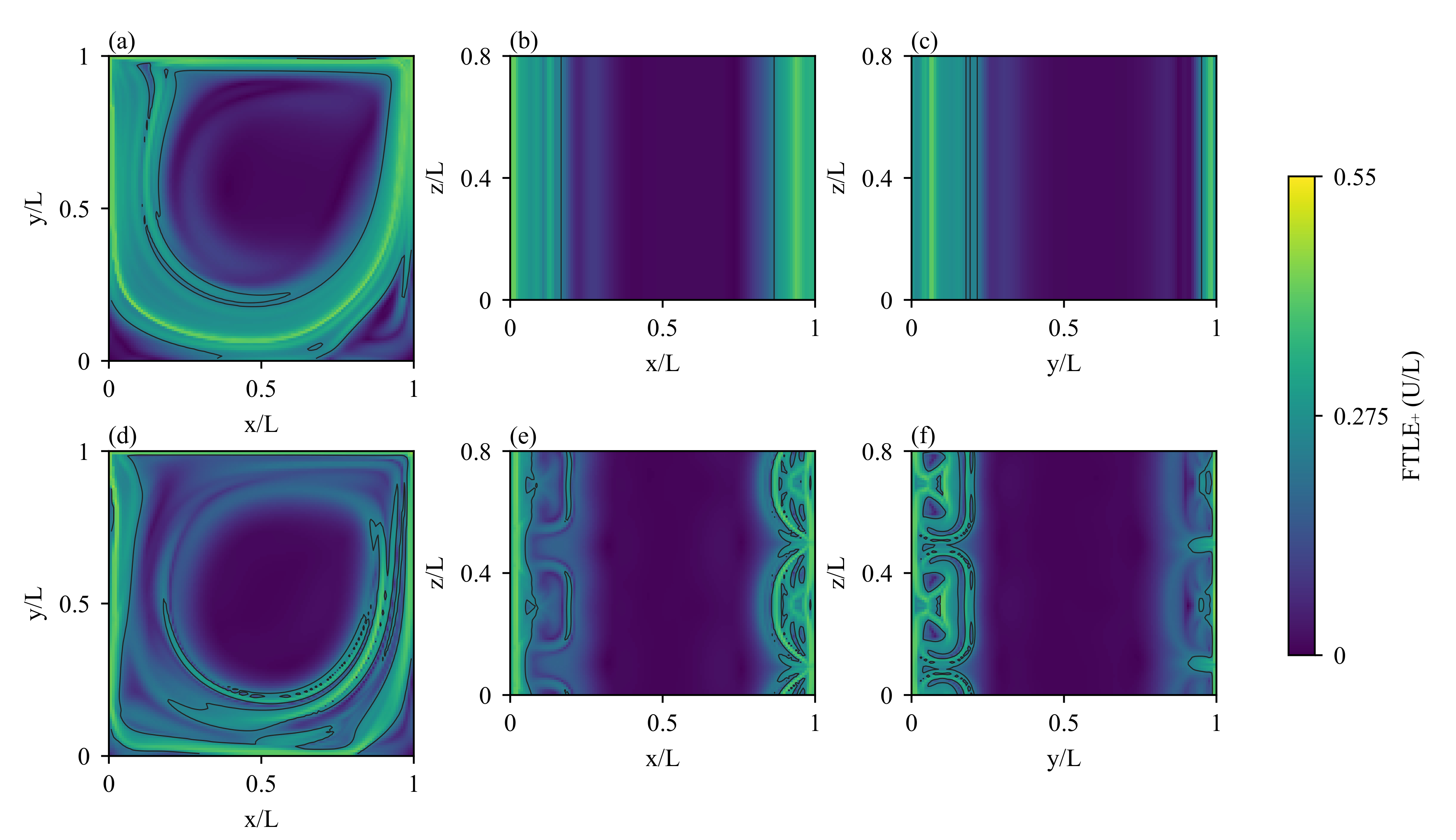}
  \caption{Forward FTLE in the periodic lid-driven cavity at $t_0=280$. (a--c) Newtonian fluid. (d--f) Carreau fluid. (a,d) the $xy$ plane at $z/L=0.4$, (b,e) the $xz$ plane at $y/L=0.5$, and (c,f) the $yz$ plane at $x/L=0.5$. }
  \label{fig:cavity-forward-sections}
\end{figure}

\begin{figure}[htbp]
  \centering
  \includegraphics[width=\textwidth]{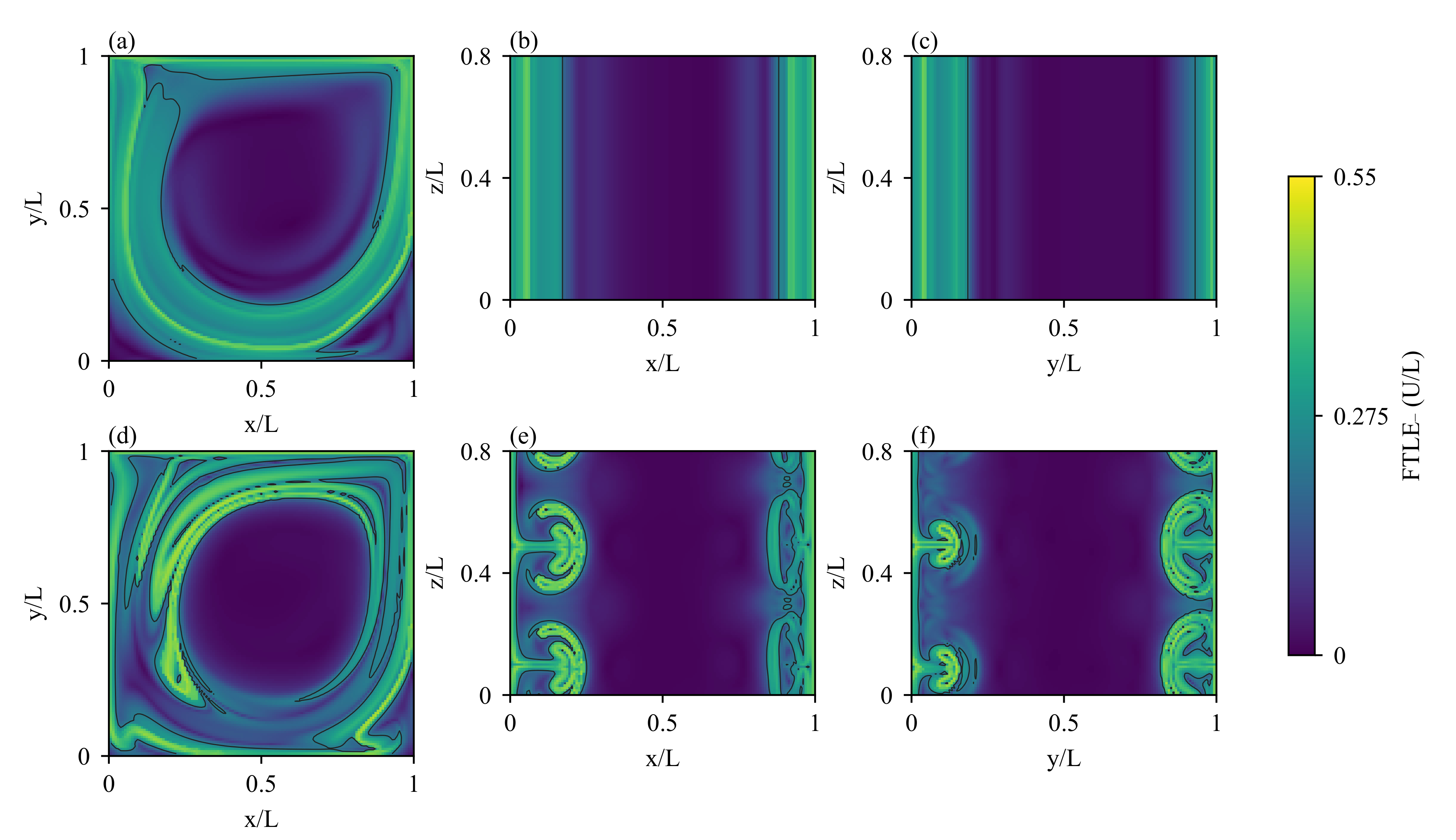}
  \caption{Backward FTLE in the periodic lid-driven cavity at $t_0=280$. (a--c) Newtonian fluid. (d--f) Carreau fluid. (a,d) the $xy$ plane at $z/L=0.4$, (b,e) the $xz$ plane at $y/L=0.5$, and (c,f) the $yz$ plane at $x/L=0.5$. }
  \label{fig:cavity-backward-sections}
\end{figure}

\begin{figure}[htbp]
  \centering
  \includegraphics[width=\textwidth]{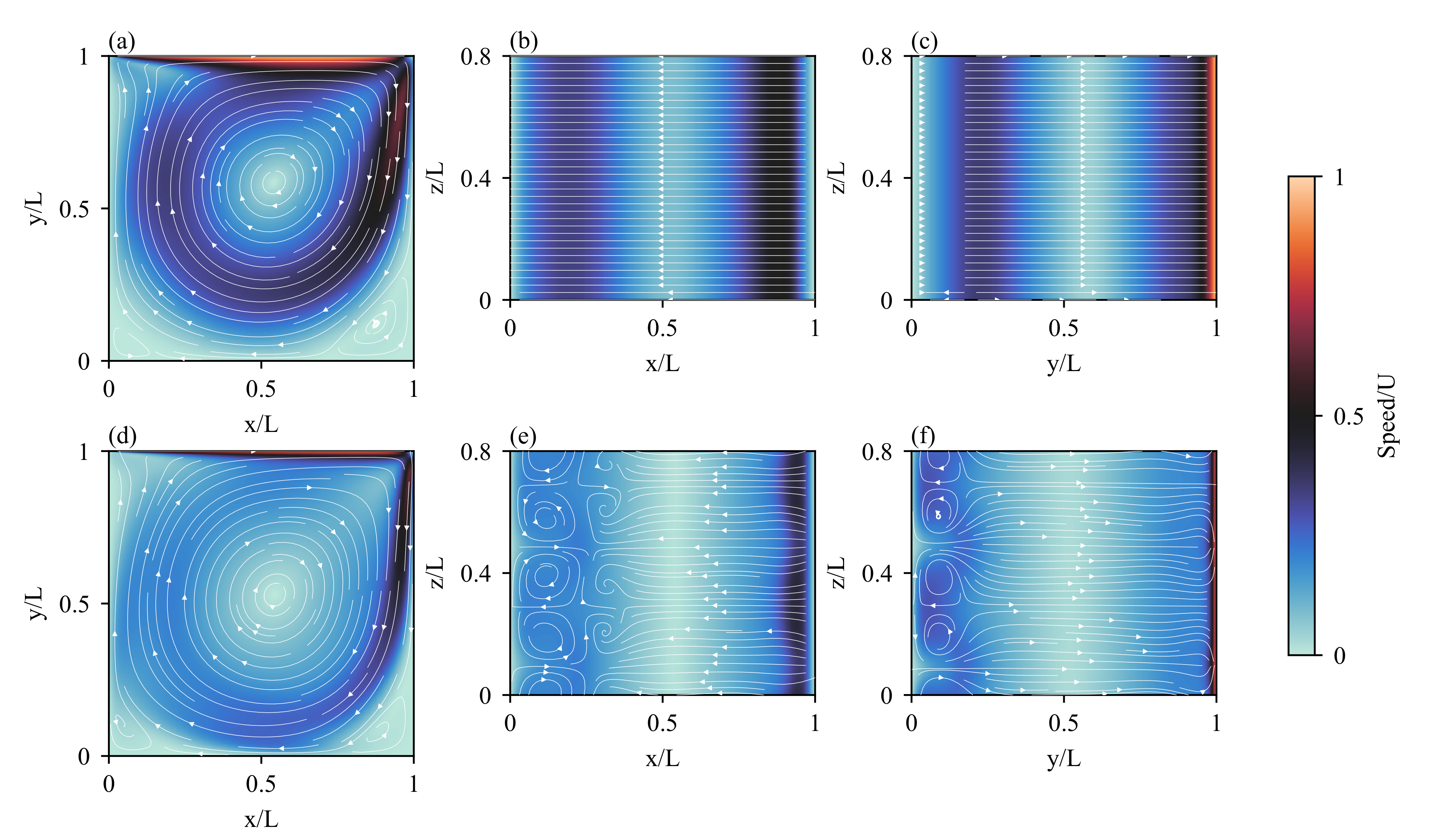}
  \caption{Instantaneous speed $|\boldsymbol{u}|/U$ at $t_0=280$. (a--c) Newtonian fluid. (d--f) Carreau fluid. (a,d) the $xy$ plane at $z/L=0.4$, (b,e) the $xz$ plane at $y/L=0.5$, and (c,f) the $yz$ plane at $x/L=0.5$.White projected streamlines follow $(u,v)$ in $xy$, $(u,w)$ in $xz$, and $(v,w)$ in $yz$.}
  \label{fig:cavity-velocity-sections}
\end{figure}

Figures~\ref{fig:cavity-forward-sections}--\ref{fig:cavity-velocity-sections} show the FTLE  and instantaneous velocity sections for the cavity flows described in \S\ref{sec:application-cavity}.

\clearpage

\begin{figure}[htbp]
  \centering
  \includegraphics[width=\textwidth]{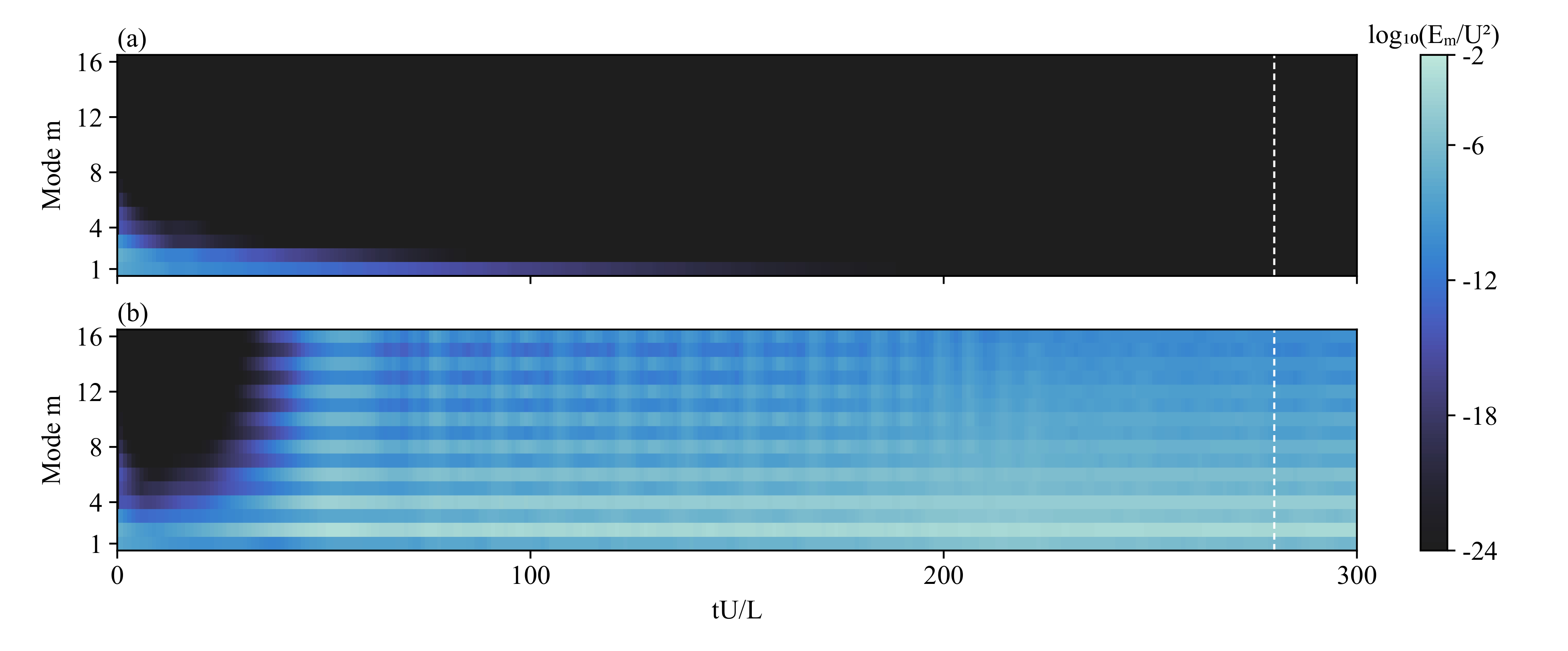}
  \caption{Spanwise Fourier-mode disturbance energy. (a) Newtonian fluid. (b) Carreau fluid.}
  \label{fig:cavity-modal-energy}
\end{figure}

For the nondimensional velocity, the energy of each conjugate spanwise Fourier pair is
\begin{equation}
  \frac{E_m(t)}{U^2}=\frac{1}{2V}\int_V
  |\boldsymbol{u}^{(m)}(\boldsymbol{x},t)|^2\,\mathrm{d}V,
  \qquad m=1,\ldots,16,
  \label{eq:cavity-modal-energy}
\end{equation}
where $\boldsymbol{u}^{(m)}$ contains modes $+m$ and $-m$, and $V=0.8$ is the nondimensional domain volume. Excluding $m=0$ gives the disturbance energy relative to the instantaneous spanwise mean. Velocity snapshots at intervals of $L/U$ are transformed in $z$ and integrated with Clenshaw--Curtis weights in $x$ and $y$. The time evolution of $E_m$ is shown in Figure~\ref{fig:cavity-modal-energy}. The Newtonian flow decays to a nearly 2-D state, while the Carreau flow develops a finite-amplitude $m=2$ mode and higher harmonics.

\clearpage

\bibliographystyle{jfm}
\bibliography{streamcenter_lcs_release}

\end{document}